\documentclass[letter,12pt]{article}
\usepackage{jheppub}
\usepackage{amsmath,amssymb,amsthm,amsfonts,mathtools}
\usepackage{bm,bbm}
\usepackage{braket}
\usepackage{cancel}
\usepackage{CJKutf8}
\usepackage{color}
\usepackage{dcolumn}
\usepackage{dsfont}
\usepackage{graphicx}
\usepackage{indentfirst}
\usepackage{mathrsfs}
\usepackage{ragged2e}
\usepackage{slashed}
\usepackage[utf8]{inputenc}
\usepackage{xcolor}
\usepackage{booktabs}
\usepackage{tabularx}
\usepackage[section]{placeins}
\usepackage[normalem]{ulem}
\definecolor{cutgray}{gray}{0.45}

\newcommand{\Tr}{\operatorname{Tr}}

\title{Electroweak Thermal Histories in the Minimal Type-II Seesaw Model with a Renormalization Group-Improved Effective Potential}

\author[a,b]{Cheng-Wei Chiang}
\emailAdd{chengwei@phys.ntu.edu.tw}
\affiliation[a]{Department of Physics and Center for Theoretical Physics,
National Taiwan University, Taipei 10617, Taiwan}
\affiliation[b]{Physics Division, National Center for Theoretical Sciences,
Taipei 10617, Taiwan}

\abstract{We study electroweak thermal histories in the minimal type-II seesaw model with a renormalization group-improved finite-temperature effective potential.  The commonly used criterion $v_c/T_c>1$, with $v_c$ the broken-phase field at the critical temperature $T_c$, applied to every first-order degeneracy of the neutral-field phase structure in our survey, does not identify a strong electroweak phase transition.  Treating those degeneracies as completed transitions would assign them a gravitational-wave signal they have not been shown to produce.  Transitions that do nucleate, percolate and complete are found for the model.  They are weak and fast, too weak on the standard sphaleron estimate to protect a baryon asymmetry from washout.  Their sound wave spectra lie far below the sensitivity of planned space detectors.  Their properties depend strongly on the renormalization group improvement and on the renormalization scale.}


\begin{document}
\maketitle
\raggedbottom

\section{Introduction}
\label{sec:introduction}

The Standard Model (SM) with the observed Higgs boson mass undergoes a smooth crossover rather than a first-order electroweak phase transition in the early Universe~\cite{Kajantie1996,Gurtler1997,Csikor1999}.  A first-order phase transition (FOPT) can instead proceed through the nucleation and growth of bubbles when the field-dependent masses of additional bosons contribute thermal terms that raise a barrier between two competing phases.  The type-II seesaw model (or Higgs triplet model) grew out of a series of early works that made use of a scalar triplet to give neutrinos mass~\cite{KonetschnyKummer1977,MaggWetterich1980,SchechterValle1980,ChengLi1980,LazaridesShafiWetterich1981,MohapatraSenjanovic1981}.  In its minimal low-energy form, the scalar sector is enlarged by one complex weak isospin triplet field of hypercharge $Y=2$.  The triplet Yukawa interaction generates Majorana neutrino masses.  The triplet scalars change the thermal history of electroweak symmetry breaking and its possible gravitational-wave (GW) signal, and the singly and doubly charged triplet states can be sought at colliders.

An early finite-temperature study of the same hypercharge-two scalar representation considered an inert triplet protected by an exact $Z_2$ symmetry on a single doublet background and with the thermal functions expanded in the mass-to-temperature ratio~\cite{AbdusSalam2014}.  Another inert triplet study retained the full thermal functions and calculated the GW spectrum, but still set the triplet vacuum expectation value (VEV) and the type-II seesaw trilinear interaction to zero~\cite{Kazemi2021}.  A complementary representation-independent analysis established how a higher scalar representation changes the sphaleron washout criterion but did not scan the thermal history of the type-II seesaw potential~\cite{Ahriche2014}.

Strong critical transitions and GW benchmarks of the minimal type-II seesaw were reported by using a fixed-scale one-loop potential~\cite{Zhou2022}.  The three benchmarks selected there were carried through to a percolation temperature, but they were excluded by the experimental searches for the doubly charged Higgs boson of the model.  Strong transitions and potentially observable GWs have also been reported after adding a singlet~\cite{Ghosh2023,Roy2025}, an inert doublet~\cite{Jangid2024}, or a gauged multitriplet sector~\cite{Park2026}.  Each added field changes the thermal potential, and these results do not apply directly to the minimal model.

These earlier studies addressed different questions, but they share one methodological issue.  An effective potential computed in a fixed gauge and truncated at a given loop order depends on the gauge choice, renormalization scale, and thermal resummation prescription~\cite{Nielsen1975,Patel2011,WainwrightProfumo2011,Croon2021}, while a physical quantity would be independent of these choices if all orders were kept.  Of the three, we study in this work the renormalization scale dependence.  The all-order potential satisfies the renormalization group (RG) equation $\mathcal{D}_{\rm RG}V_{\rm eff}=0$, where $\mathcal{D}_{\rm RG}=Q\,\partial_Q+\sum_X\beta_X\,\partial_X-\gamma_\Phi\,\varphi\,\partial_\varphi-\gamma_\Delta\,\delta\,\partial_\delta$ collects the explicit dependence on the renormalization scale $Q$ together with the running of every Lagrangian parameter $X$ and of both classical fields, the neutral doublet and triplet backgrounds $\varphi$ and $\delta$, with $\beta_X=dX/d\log Q$ and $\gamma_\Phi$, $\gamma_\Delta$ the anomalous dimensions of the two fields.  A potential truncated at $\ell$ loops leaves $\mathcal{D}_{\rm RG}V_{\rm eff}^{(\ell)}=\mathcal{O}(\hbar^{\ell+1})$.  The residual measures how sensitive a result is to the omitted orders.  A critical FOPT, a solution of the bounce equations, nucleation, percolation, and completion are distinct issues in the cosmological evolution.  None of the studies above tests all of them within an RG-improved treatment of the minimal type-II seesaw model.  A treatment of this kind, with a thermally resummed one-loop potential whose couplings and classical fields are both evolved by the RG equations, has recently been applied to the Georgi--Machacek model~\cite{ChiangEnomoto2026}, which adds a real triplet to the same scalar sector.

The RG equations of the minimal type-II seesaw model are already well established.  The complete one-loop evolution of the Higgs and lepton couplings was derived in Ref.~\cite{ChaoZhang2007} and independently analyzed in Ref.~\cite{Schmidt2007}.  Early applications to Higgs mass bounds and large-scale vacuum stability include Refs.~\cite{Gogoladze2008,ChaoGonderinger2012,Chun2012,Dev2013}.  Later analyses incorporated threshold matching and, in some cases, higher-order running~\cite{Kobakhidze2013,Haba2016,GhoshHighScale2018}.  The necessary and sufficient conditions for the potential to be bounded from below, including correlations among the gauge orbit variables that were missed in some earlier treatments, were clarified in Ref.~\cite{Bonilla2015}.  In our study, we use the established complete scalar spectrum~\cite{Arhrib2011} and the exact correlated boundedness conditions~\cite{Bonilla2015}, construct a one-loop RG-improved effective potential with a field- and temperature-dependent scale, and vary the renormalization scale at every completed point.  The calculation uses full thermal functions and Arnold--Espinosa resummation~\cite{ArnoldEspinosa1993}, supplemented by a mass-dependent check of the high-temperature screening approximation.  We use CosmoTransitions (CT)~\cite{Wainwright2012} to trace the complete neutral-field phase history, and solve the bounce equation between the two minima for the tunneling action.  A transition is conventionally taken to occur once the three-dimensional bounce action reaches $S_3/T\simeq140$, the value at which about one bubble is nucleated per Hubble volume per Hubble time at the electroweak scale.  Before calculating GW spectra, we distinguish this estimate from integrated nucleation, percolation, and completion.

We first perform a broad survey of points that satisfy the zero-temperature requirements to examine what is inferred from a critical-temperature criterion alone.  A condition such as $v_c/T_c>1$, with $v_c$ the broken-phase field at the critical temperature $T_c$ at which two phases are degenerate, applied without checking which minima the two phases correspond to, admits a small fraction of the resolved first-order critical transitions in the survey as apparently strong.  It does not distinguish one-step from multistep histories.  Those candidates carry transition strengths $5$ to $7$ orders of magnitude above the strongest transition that completes eventually.  Through the known scaling of the acoustic amplitude, this is at least $6$ orders of magnitude in the strength-dependent factor of the signal.  This hierarchy shows how an analysis based only on the critical temperature can produce an apparently large GW signal.  Nucleation, percolation, and completion must all be established before interpreting such a spectrum as a cosmological prediction.

Our subsequent search finds seven benchmark points whose transitions successfully complete and which remain consistent with the electroweak fit and with the collider limits that apply at their triplet VEVs.  We identify these points by their Lagrangian parameters and draw no conclusion from their relative placement.  Conditional on the plasma reaching the symmetric phase, every retained channel in this completed sample is a direct transition from it to the electroweak vacuum.  The transition is weak in the central prescription, too weak for the broken-phase sphaleron rate to fall below the Hubble rate at nucleation.  The resulting GW spectra are at least $10$ orders of magnitude below the sensitivities of the Laser Interferometer Space Antenna (LISA), Deci-hertz Interferometer Gravitational-wave Observatory (DECIGO), and Big Bang Observer (BBO) experiments.

The rest of this paper is organized as follows.  Section~\ref{sec:model-constraints} presents the model, its scalar spectrum, and the zero-temperature theoretical and experimental constraints.  Section~\ref{sec:one-loop-rg} constructs the one-loop effective potential and its RG improvement, and Sec.~\ref{sec:thermal-potential} adds the thermal corrections.  Section~\ref{sec:ct-methods} describes the thermal histories, tunneling actions, cosmological criteria, and GW spectra.  Section~\ref{sec:scan-results} shows what an analysis based on the critical temperature alone misses, and Sec.~\ref{sec:completed-transition-benchmarks} analyzes the seven completed transitions, their thermal properties, scale dependence, and GW signals.  Section~\ref{sec:conclusions} summarizes our findings.

\section{The Type-II Seesaw Model}
\label{sec:model-constraints}

\subsection{Field content, inputs, and tree-level vacuum}
\label{subsec:field-content}

This model extends the SM Higgs sector with a complex triplet that carries the hypercharge $Y=2$ (in the convention of the electric charge $Q_{\rm em}=T_3+Y/2$).  We assume a potential with real couplings that conserves the combined charge conjugation and parity transformation ($CP$), together with a real triplet VEV.  Our reference scale is the electroweak scale $Q_{\rm ref} = v\equiv \sqrt{v_\Phi^2+2v_\Delta^2}$, with $v_\Phi$ and $v_\Delta$ denoting the doublet and triplet VEVs, respectively.  Numerically, $v=246.2$~GeV is fixed by the Fermi constant at tree level~\cite{PDG2026}.  The mass of the Higgs boson $h$ is set to the Particle Data Group (PDG) value $m_h=125.13$~GeV~\cite{PDG2026}.  At this reference scale, we use $(g_2,g')=(0.6462,0.3594)$, $g_3=1.1412$, and $y_t=0.9155$ as the running inputs for the electroweak, strong, and top Yukawa couplings, respectively.  Each is derived from measurements: the two electroweak couplings from the running electromagnetic coupling and the modified minimal subtraction ($\overline{\mathrm{MS}}$) weak mixing angle at the $Z$ boson mass $M_Z$, and $g_3$ from the strong coupling $\alpha_s(M_Z)$, each taken from Ref.~\cite{PDG2026} and then evolved to $Q_{\rm ref}$ at one loop.

The top Yukawa coupling is the one input whose value depends on the top mass scheme.  We take the $\overline{\mathrm{MS}}$ coupling at $Q_{\rm ref}$, obtained from the direct measurement average $M_t=172.60$~GeV of Ref.~\cite{PDG2026}, converted with the three-loop quantum chromodynamics series to $m_t(M_t)\simeq162.3$~GeV and evolved with the one-loop SM Yukawa equation to $Q_{\rm ref}$, which gives the value $0.9155$ quoted above.  The pole mass used directly in $y_t=\sqrt{2}\,M_t/v$ would give a value larger by about $8\%$, and is the wrong scheme for a potential using the $\overline{\mathrm{MS}}$ scheme.

The scalar fields are parametrized as
\begin{equation}
 \Phi=\begin{pmatrix}\phi^+\\(v_\Phi+h_\Phi+i\chi)/\sqrt2\end{pmatrix},\qquad
 \Delta=\begin{pmatrix}
 \Delta^+/\sqrt2&H^{++}\\(v_\Delta+h_\Delta+i\eta)/\sqrt2&-\Delta^+/\sqrt2
 \end{pmatrix}
 ~.
\end{equation}
The most general renormalizable potential that conserves $CP$ is then given by\footnote{Lagrangians with different forms of operators and coupling names appear in the literature.  For a traceless $2\times2$ triplet matrix, we have
\begin{equation}
 \Delta\Delta^\dagger+\Delta^\dagger\Delta
 =\Tr(\Delta^\dagger\Delta)\,\mathbbm{1}
 ~, \qquad
 \Phi^\dagger\Delta^\dagger\Delta\Phi
 =\Tr(\Delta^\dagger\Delta)\Phi^\dagger\Phi
 -\Phi^\dagger\Delta\Delta^\dagger\Phi
 ~.
 \label{eq:portal-identity}
\end{equation}
Thus, of the three doublet--triplet quartics $(\Phi^\dagger\Phi)\Tr(\Delta^\dagger\Delta)$, $\Phi^\dagger\Delta\Delta^\dagger\Phi$ and $\Phi^\dagger\Delta^\dagger\Delta\Phi$, only two are independent, and we keep the first two.  Equation~\eqref{eq:portal-identity} also gives $\Phi^\dagger[\Delta,\Delta^\dagger]\Phi=2\Phi^\dagger\Delta\Delta^\dagger\Phi-\Tr(\Delta^\dagger\Delta)\Phi^\dagger\Phi$, so a term $c\,\Phi^\dagger[\Delta,\Delta^\dagger]\Phi$ contributes $-c$ and $2c$ to our $\lambda_4$ and $\lambda_5$, respectively.  }
\begin{align}
 V={}&-m^2\Phi^\dagger\Phi+M^2\Tr(\Delta^\dagger\Delta)
 +\left[\mu\Phi^{\top}i\tau_2\Delta^\dagger\Phi+\mathrm{H.c.}\right]
 +\lambda_1(\Phi^\dagger\Phi)^2 \nonumber\\
 &+\lambda_2[\Tr(\Delta^\dagger\Delta)]^2
 +\lambda_3\Tr(\Delta^\dagger\Delta\Delta^\dagger\Delta)
 +\lambda_4(\Phi^\dagger\Phi)\Tr(\Delta^\dagger\Delta)
 +\lambda_5\Phi^\dagger\Delta\Delta^\dagger\Phi
 ~.
 \label{eq:potential}
\end{align}
The Pauli matrix $\tau_2$ acts on the weak isospin indices, and H.c.\ denotes the Hermitian conjugate.  With the convention $L(\Phi)=0$ and $L(\Delta)=-2$, the term proportional to the dimensionful parameter $\mu$ breaks lepton number by two units.

We neglect the triplet--lepton Yukawa matrix $Y_\Delta$ in the effective potential and in the running.  Its entries scale as $m_\nu/v_\Delta$ and stay far below the gauge and quartic couplings over the triplet VEV range used here, $1.0\times10^{-5}\leq v_\Delta/\mathrm{GeV} \leq 2$.  Their one-loop contributions to the scalar beta functions are negligible at this accuracy~\cite{ChaoZhang2007,Schmidt2007}.

On the real neutral backgrounds $\Phi^0=\varphi/\sqrt2$ and $\Delta^0=\delta/\sqrt2$, with all charged fields set to zero, the potential becomes
\begin{equation}
 V_0=-\frac12m^2\varphi^2+\frac12M^2\delta^2
 -\frac{\mu}{\sqrt2}\varphi^2\delta
 +\frac14\lambda_1\varphi^4
 +\frac14(\lambda_2+\lambda_3)\delta^4
 +\frac14(\lambda_4+\lambda_5)\varphi^2\delta^2
 ~.
 \label{eq:neutralpotential}
\end{equation}
Setting $\partial V_0/\partial\varphi=\partial V_0/\partial\delta=0$ at $(\varphi,\delta)=(v_\Phi,v_\Delta)$ gives the tadpole conditions
\begin{subequations}
\label{eq:tadpoles}
\begin{align}
 m^2&=\lambda_1v_\Phi^2+\frac12(\lambda_4+\lambda_5)v_\Delta^2
 -\sqrt2\mu v_\Delta
 ~,
 \label{eq:tadpole-doublet}\\
 \frac{\mu v_\Phi^2}{\sqrt2v_\Delta}
 &=M^2+(\lambda_2+\lambda_3)v_\Delta^2
 +\frac12(\lambda_4+\lambda_5)v_\Phi^2
 ~.
 \label{eq:tadpole-triplet}
\end{align}
\end{subequations}
We define, with $\lambda_{ij} \equiv \lambda_i+\lambda_j$,
\begin{equation}
 M_\Delta^2\equiv\frac{\mu v_\Phi^2}{\sqrt2v_\Delta}
 =M^2+\lambda_{23}v_\Delta^2+\frac12\lambda_{45}v_\Phi^2
 ~.
 \label{eq:MDelta-distinction}
\end{equation}

When $M^2$ dominates the field-dependent correction to the triplet curvature, the triplet can be eliminated perturbatively.  On the branch connected continuously to $\delta=0$, its tree-level equation of motion gives $\delta(\varphi)=\mu\varphi^2/[\sqrt2(M^2+\lambda_{45}\varphi^2/2)]+\mathcal{O}(\varphi^6)$ once $\lambda_{23}\delta^3$ is dropped.  Substituting it back into $V_0$ gives the part of the reduced doublet potential induced by the triplet,
\begin{align}
 \Delta V_0^{(\delta)}
 &=-\frac{\mu^2\varphi^4}
 {4\left(M^2+\lambda_{45}\varphi^2/2\right)}
 +\mathcal{O}(\varphi^8)
\simeq -\frac{\mu^2}{4M^2}\varphi^4
 +\frac{\lambda_{45}\mu^2}{8M^4}\varphi^6
~.
\label{eq:expansion}
\end{align}
Comparing the reduced potential with the convention $(\Delta\lambda_1/4)\varphi^4+(c_6/8\Lambda^2)\varphi^6$ gives
\begin{equation}
 \Delta\lambda_1=-\frac{\mu^2}{M^2}
 ~,\qquad
 \frac{c_6}{\Lambda^2}=\frac{\lambda_{45}\mu^2}{M^4}
 ~,\qquad
 \Lambda=\frac{M^2}{|\mu|\sqrt{|\lambda_{45}|}}
 ~,\qquad
 c_6\equiv\operatorname{sgn}(\lambda_{45})
 ~.
\end{equation}
We fix $|c_6|=1$, so that $\Lambda$ carries the entire magnitude and the sign of $c_6/\Lambda^2$ is the sign of $\lambda_{45}$.  Equation~\eqref{eq:expansion} holds where $|\lambda_{45}|\varphi^2/(2M^2)\ll1$, which controls the denominator, and $|\lambda_{23}|\mu^2\varphi^4/(2M^6)\ll1$, which controls the cubic term dropped above.  Within that domain, the induced sextic term can stabilize a negative doublet quartic only for $\lambda_{45}>0$.  Stationarity of the reduced potential $V_{\rm red}$ at $\varphi=v_\Phi$ together with $V_{\rm red}''(v_\Phi)=m_h^2$ gives $m_h^2=2\lambda_{\rm eff}v_\Phi^2+3(c_6/\Lambda^2)v_\Phi^4$, where $\lambda_{\rm eff}=\lambda_1+\Delta\lambda_1$ is the doublet quartic after the triplet has been integrated out.  Hence, stabilizing a negative doublet quartic requires $c_6/\Lambda^2>m_h^2/(3v_\Phi^4)$.  Eliminating $\mu$ with Eq.~\eqref{eq:MDelta-distinction}, $\mu=\sqrt2\,v_\Delta M_\Delta^2/v_\Phi^2$, turns this into $2\lambda_{45}v_\Delta^2M_\Delta^4/M^4>m_h^2/3$, that is
\begin{equation}
 v_\Delta\sqrt{\lambda_{45}}\,\frac{M_\Delta^2}{M^2}\gtrsim\frac{m_h}{\sqrt6}
 ~.
 \label{eq:tree-sextic-bound}
\end{equation}
Inside the domain where Eq.~\eqref{eq:expansion} holds, $M_\Delta^2/M^2=1+\lambda_{45}v_\Phi^2/(2M^2)+\lambda_{23}v_\Delta^2/M^2$ differs from one only by the small quantities that control the expansion, so the bound reads $v_\Delta\sqrt{\lambda_{45}}\gtrsim m_h/\sqrt6$ there.  With $|\lambda_{45}|$ bounded by perturbative unitarity, $|\lambda_{45}|\leq8\pi$, and triplet VEVs below $2$~GeV, the left-hand side of Eq.~\eqref{eq:tree-sextic-bound} stays far below $m_h/\sqrt6$, so the bound is never satisfied here.  This rules out a barrier from the induced doublet sextic term.  A tree-level barrier involving both $\varphi$ and $\delta$ remains possible, and the full finite-temperature potential decides where any barrier comes from.

\subsection{Scalar spectrum}
\label{subsec:spectrum}

Expanding Eq.~\eqref{eq:potential} in the real field components and evaluating the Hessian at the vacuum $(v_\Phi,v_\Delta)$ after imposing Eq.~\eqref{eq:tadpoles} gives three massless modes, one for each broken generator of $SU(2)_L\times U(1)_Y$.  The physical charged and $CP$-odd masses are
\begin{align}
 m_{H^{\pm\pm}}^2&=M_\Delta^2-\lambda_3v_\Delta^2
 -\frac12\lambda_5v_\Phi^2
 ~,\\
 m_{H^\pm}^2&=\left(M_\Delta^2-\frac14\lambda_5v_\Phi^2\right)
 \left(1+\frac{2v_\Delta^2}{v_\Phi^2}\right)
 ~,\\
 m_A^2&=M_\Delta^2\left(1+\frac{4v_\Delta^2}{v_\Phi^2}\right)
 ~.
\end{align}
The squared mass matrix in the $CP$-even gauge basis $(h_\Phi,h_\Delta)$ is
\begin{equation}
 \mathcal{M}_{\rm even}^2=\begin{pmatrix}
 2\lambda_1v_\Phi^2&(\lambda_4+\lambda_5)v_\Phi v_\Delta
 -2M_\Delta^2v_\Delta/v_\Phi\\
 (\lambda_4+\lambda_5)v_\Phi v_\Delta-2M_\Delta^2v_\Delta/v_\Phi&
 M_\Delta^2+2(\lambda_2+\lambda_3)v_\Delta^2
 \end{pmatrix}
 ~.
\end{equation}
We denote its eigenstates by $h$ and $H^0$, with $h$ the state of mass $m_h$, which is mostly the doublet.  They are given by $h=\cos\alpha\,h_\Phi+\sin\alpha\,h_\Delta$ and $H^0=-\sin\alpha\,h_\Phi+\cos\alpha\,h_\Delta$, and the mixing angle obeys
\begin{equation}
 \tan2\alpha=
 \frac{2(\mathcal{M}_{\rm even}^2)_{12}}
 {(\mathcal{M}_{\rm even}^2)_{11}-(\mathcal{M}_{\rm even}^2)_{22}}
 ~.
\end{equation}
Requiring the eigenvalue of $h$ to equal $m_h^2$ fixes $\lambda_1$ through the full $CP$-even matrix above, not through its $(1,1)$ entry alone.  We use this tree-level determination at every parameter point.

The angle $\beta_\pm$ rotates $(\phi^\pm,\Delta^\pm)$ into the charged Goldstone mode and the physical $H^\pm$, and $\beta_0$ rotates $(\chi,\eta)$ into the neutral Goldstone mode and the physical $A$, while the doubly charged state does not mix.  Both angles are fixed by the vacuum,
\begin{equation}
 \tan\beta_\pm=\frac{\sqrt2v_\Delta}{v_\Phi}
 ~,\qquad
 \tan\beta_0=\frac{2v_\Delta}{v_\Phi}
 ~.
\end{equation}

For $v_\Delta\ll v_\Phi$, as considered here, the three mostly triplet states $H^{\pm\pm}$, $H^\pm$, and $A$ satisfy $m_{H^{\pm\pm}}^2-m_{H^\pm}^2\simeq m_{H^\pm}^2-m_A^2\simeq-\lambda_5v_\Phi^2/4$, so they are approximately equally spaced in squared mass and the sign of $\lambda_5$ decides whether the doubly charged state is the lightest or the heaviest~\cite{Arhrib2011,Aoki2012MassDifference,Kanemura2012}.  The same $\lambda_5$ fixes which decay chains among the triplet states are open, and it also enters the thermal masses of both multiplets (Sec.~\ref{subsec:self-energies}).  The mass ordering alone does not imply a FOPT.

\subsection{Exact boundedness from below}
\label{subsec:bfb}

At large field values, we define
\begin{equation}
 r=\Phi^\dagger\Phi,\quad t=\Tr(\Delta^\dagger\Delta)
 ~,\quad
 \zeta=\frac{\Tr(\Delta^\dagger\Delta\Delta^\dagger\Delta)}{t^2}
 ~,\quad
 \xi=\frac{\Phi^\dagger\Delta\Delta^\dagger\Phi}{rt}
 ~,
\end{equation}
and consider $r>0$ and $t>0$.  The quartic potential is then given by
\begin{equation}
 V_4=\lambda_1r^2+(\lambda_4+\lambda_5\xi)rt
 +(\lambda_2+\lambda_3\zeta)t^2
 ~.
 \label{eq:v4}
\end{equation}
Separately, $1/2\leq\zeta\leq1$ and $0\leq\xi\leq1$.  These intervals are not independent.  Let $st$ and $(1-s)t$, with $0\leq s\leq1$, be the eigenvalues of $\Delta^\dagger\Delta$, and let $u$, with $0\leq u\leq1$, be the fraction of $r$ aligned with the eigenvector of $\Delta\Delta^\dagger$ belonging to $st$.  Then
\begin{equation}
 \zeta=s^2+(1-s)^2
 ~,\qquad
 \xi=us+(1-u)(1-s)
 ~,\qquad
 (2\xi-1)^2\leq2\zeta-1
 ~.
 \label{eq:orbit}
\end{equation}
The last relation follows because the first two give $2\zeta-1=(2s-1)^2$ and $2\xi-1=(2u-1)(2s-1)$, with $|2u-1|\leq1$.  The directions $r=0$ and $t=0$ are evaluated directly from $V_4$.  Equation~\eqref{eq:orbit} characterizes exactly the region of the $(\zeta,\xi)$ plane that field configurations can reach once gauge-equivalent ones are identified.  Requiring $V_4\geq0$ over the entire rectangle $1/2\leq\zeta\leq1$, $0\leq\xi\leq1$ instead of over the above region is therefore sufficient but not necessary, and can reject a stable quartic potential.  The separate ranges of $\zeta$ and $\xi$ quoted in Ref.~\cite{Arhrib2011} do not encode their correlation.  We use the correlated boundedness conditions established in Ref.~\cite{Bonilla2015}.  Satisfied as strict inequalities, they make $V_4$ positive for every nonzero field and so bound the potential from below.  When one of them holds as an equality, $V_4$ vanishes along some field direction, and the cubic and quadratic terms decide along it.

The pure doublet direction $t=0$ and the pure triplet direction $r=0$ give
\begin{equation}
 \lambda_1\geq0
 ~,\qquad
 \lambda_2+\lambda_3\geq0
 ~,\qquad
 2\lambda_2+\lambda_3\geq0
 ~.
\end{equation}
With $w=|2s-1|$, so that $\zeta=(1+w^2)/2$, minimizing over the doublet orientation gives $\min_u(\lambda_5\xi)=\lambda_5/2-|\lambda_5|w/2$.  Applying the two-variable copositivity condition to the resulting quadratic form in $r$ and $t$ at every $0\leq w\leq1$ gives the mixed condition
\begin{equation}
 \min_{0\leq w\leq1}\left[
 \lambda_4+\frac{\lambda_5}{2}-\frac{|\lambda_5|w}{2}
 +\sqrt{2\lambda_1(2\lambda_2+\lambda_3+\lambda_3w^2)}
 \right]\geq0
 ~.
 \label{eq:exactbfb}
\end{equation}
The minimum in the last condition can lie inside the interval.  So endpoint checks are insufficient.  A test at the endpoints alone could accept a quartic potential that is unbounded from below, which is why Eq.~\eqref{eq:exactbfb} is imposed as a minimization.

\subsection{Perturbative unitarity and individual coupling bounds}
\label{subsec:unitarity}

High-energy two-scalar scattering imposes $|\operatorname{Re}a_0|\leq1/2$ on every eigenvalue of the zeroth partial-wave amplitude.  In the convention of Eq.~\eqref{eq:potential}, the twelve conditions may be written as~\cite{Arhrib2011}
\begin{align}
\begin{split}
 |\lambda_1|\leq4\pi
 ~,\quad
 |\lambda_2|\leq4\pi
 ~,\quad
 |\lambda_{23}|\leq4\pi
 ~,\quad
 |\lambda_4|\leq8\pi
 ~,\quad
 |\lambda_{45}|\leq8\pi
 ~, \\
 |2\lambda_2-\lambda_3|\leq8\pi
 ~,\quad
 \left|\lambda_4-\frac{\lambda_5}{2}\right|\leq8\pi
 ~,\quad
 |2\lambda_4+3\lambda_5|\leq16\pi
 ~, \\
 \left|\lambda_{12}+2\lambda_3
 \pm\sqrt{(\lambda_1-\lambda_2-2\lambda_3)^2+\lambda_5^2}\right|
 \leq8\pi
 ~, \\
 \left|3\lambda_{13}+4\lambda_2
 \pm\sqrt{(3\lambda_1-4\lambda_2-3\lambda_3)^2
 +\frac32(2\lambda_4+\lambda_5)^2}\right|\leq8\pi
 ~.
\end{split}
 \label{eq:tree-unitarity}
\end{align}
The individual perturbativity bound $|\lambda_i|\leq4\pi$ is imposed separately.  Equation~\eqref{eq:tree-unitarity} constrains scattering eigenchannels, not each Lagrangian coefficient one by one, and does not imply the individual bound for the doublet--triplet quartic couplings $\lambda_4$ and $\lambda_5$.  A point can pass tree-level unitarity while containing one of these quartic couplings too large for a one-loop thermal calculation to be credible.  We impose the exact boundedness conditions, all twelve unitarity conditions, and the separate individual coupling bound at the reference scale $Q_{\rm ref}$.  Along the RG interval used by the effective potential, we also require every running dimensionless coupling $X$, the gauge and top Yukawa couplings included, to satisfy $|X(Q)|\leq4\pi$.  The boundedness and unitarity conditions are not reimposed at every intermediate scale.

\subsection{Collider and electroweak constraints}
\label{subsec:collider-constraints}

The triplet VEV modifies the electroweak $\rho$ parameter already at tree level, and expanding in $v_\Delta^2/v^2\ll1$ at fixed $v$ gives
\begin{equation}
 \rho_{\rm tree}=
 \frac{v_\Phi^2+2v_\Delta^2}{v_\Phi^2+4v_\Delta^2}
 \simeq1-\frac{2v_\Delta^2}{v^2}
 ~,
 \label{eq:rho-tree}
\end{equation}
so electroweak data constrain $v_\Delta$.  Scalar mass splittings also contribute at one loop to the oblique parameters $S,T,U$, which parametrize new physics contributions to the electroweak gauge boson self-energies.  The $T$ parameter receives both the tree-level shift of Eq.~\eqref{eq:rho-tree} and a one-loop contribution from the mass splittings.  Over the points that pass every condition of this section, the tree-level shift never exceeds $0.017$ in $T$, about one seventh of the experimental uncertainty $\sigma_T=0.12$, and is far smaller at most retained points.  It matters only at the boundary of the allowed region, where setting it to zero removes a few points.  We use the general scalar multiplet result of Ref.~\cite{LavouraLi1994} in the Higgs triplet specialization of Ref.~\cite{Chun2012}, with the second loop function of $U$ in the corrected form of Ref.~\cite{Ashanujjaman2026}.  The definitions follow those of Ref.~\cite{PDG2026}, whose fit subtracts the photon self-energies at $M_Z^2$ rather than at zero momentum.  With $U$ an independent parameter, the PDG fit~\cite{PDG2026} gives $S=0.018\pm0.095$, $T=0.04\pm0.12$ and $U=-0.013\pm0.086$, with a correlation matrix
\begin{equation}
 C_{STU}=
 \begin{pmatrix}
 1 & 0.91 & -0.66 \\
 0.91 & 1 & -0.88 \\
 -0.66 & -0.88 & 1
 \end{pmatrix}
 ~.
\end{equation}
With $X=(S,T,U)^{\top}$, the central values $X_0$ and standard deviations $\sigma_i$ quoted above, and the covariance matrix $\Sigma_{ij}=\sigma_i(C_{STU})_{ij}\sigma_j$, we retain points with $\chi^2=(X-X_0)^{\top}\Sigma^{-1}(X-X_0)\leq7.815$, the 95th percentile of the $\chi^2$ distribution with three degrees of freedom.  The condition is not symmetric under a flip of the sign of the triplet mass splittings.  In Ref.~\cite{LavouraLi1994}, $M_I$ denotes the squared mass $m_I^2$ of the multiplet component with third isospin component $I$.  Writing $m_I^2=\mathcal{M}^2+\Delta_I$, with $\mathcal{M}^2$ the common squared mass, $\Delta_I$ the squared mass splittings, and $s_W$ and $c_W$ the sine and cosine of the weak mixing angle, we have $T=[16\pi s_W^2c_W^2M_Z^2]^{-1}\sum_I(J^2-I^2+J+I)\,\theta_+(m_I^2,m_{I-1}^2)$, with $J$ the weak isospin of the multiplet and $\theta_+(m_I^2,m_K^2)=(\Delta_I-\Delta_K)^2/(3\mathcal{M}^2)+\mathcal{O}(\Delta_I^3/\mathcal{M}^4)$.  At leading order, $T$ is \emph{quadratic and even} in the splittings.  The splitting-dependent part of $S$ is $-(Y/6\pi)\sum_II\log(m_I^2/\bar\mu^2)$, whose dependence on the arbitrary scale $\bar\mu$ cancels through the sum rule $\sum_II=0$.  Expanding $\log(m_I^2/\bar\mu^2)=\log(\mathcal{M}^2/\bar\mu^2)+\Delta_I/\mathcal{M}^2+\mathcal{O}(\Delta_I^2/\mathcal{M}^4)$ and using the same sum rule, it is $-(1/3\pi)\sum_II\,\Delta_I/\mathcal{M}^2$ for the triplet, \emph{linear and odd} in the splittings.  Both statements hold at leading order in the $\Delta_I/\mathcal{M}^2$ expansion.  Because $S$ is odd in the splittings and correlated with $T$ in the fit, the two mass orderings fare differently.  The largest accepted $T$ is about $0.11$ for $\lambda_5>0$ and $0.07$ for $\lambda_5<0$.  As a preselection condition, we include the tree-level contribution of the triplet VEV and the one-loop contribution of the mass splittings, without a complete electroweak fit with on-shell masses.

The phenomenology of the doubly charged Higgs boson depends strongly on $v_\Delta$.  Its leading leptonic and vector boson widths scale oppositely with $v_\Delta$~\cite{FileviezPerez2008}:
\begin{align}
 \Gamma_{\ell\ell}
 &=\frac{m_{H^{\pm\pm}}}{16\pi v_\Delta^2}
   \sum_i m_{\nu_i}^2
   ~, \\
 \Gamma_{WW}
 &=\frac{g_2^4v_\Delta^2\beta_W}
 {64\pi m_{H^{\pm\pm}}x_W^2}
 \left[8x_W^2+(1-2x_W)^2\right]
 ~,
\end{align}
where $ x_W = m_W^2 / m_{H^{\pm\pm}}^2$, $\beta_W = \sqrt{1-4x_W}$ when both $W$ bosons are on shell, and $m_{\nu_i}$ are the light neutrino masses.  We take them in the normal ordering with a massless lightest state and the NuFIT~6.0 oscillation parameters~\cite{NuFIT2024}, which enter the calculation only through the decay widths of the triplet states.  The triplet Yukawa coupling, neglected in the effective potential and in the running, is kept in these widths.  Let $v_\Delta^{\rm eq}$ be the triplet VEV at which the two widths are equal at fixed $m_{H^{\pm\pm}}$.  The same-sign dilepton channels dominate for $v_\Delta\ll v_\Delta^{\rm eq}$ and the same-sign diboson channel for $v_\Delta\gg v_\Delta^{\rm eq}$, while near $v_\Delta^{\rm eq}$ the two are comparable.  Wherever the dilepton channels are open, the neutrino spectrum and Majorana phases fix their lepton flavor pattern~\cite{ChunLeePark2003,FileviezPerez2008,AkeroydChiang2009}.  Vector bosons off their mass shell and the cascade $H^{++}\to H^+W^{+*}$ can further change the mixed same-sign signature~\cite{FileviezPerez2008,Chiang2026Mixed}.  Where they are negligible, the sum of the two widths sets the proper decay length.  Where that length is macroscopic, a search assuming a prompt decay does not apply and one should resort to a displaced-vertex search.

We therefore apply each direct-search limit only where its stated decay, lifetime, and spectrum assumptions are satisfied.  The low-mass reference limits include the 43~GeV constraint that Ref.~\cite{KanemuraYagyuYokoya2013} derives from the $Z$ width, the 99.0~GeV bound set at LEP by the OPAL Collaboration on the $T_3=\pm1$ state in the $\tau\tau$ channel~\cite{OPAL2002}, the mode-dependent 210 to 245~GeV bounds of the CDF Collaboration for the $T_3=\pm1$ state~\cite{CDF2011}, and the approximate 84~GeV same-sign $WW$ reinterpretation of an 8~TeV search~\cite{ATLAS2015SameSign,Kanemura2015}.  The 139~fb$^{-1}$ dilepton search at the Large Hadron Collider (LHC) by the ATLAS experiment reports $1080$~GeV for the left--right symmetric type-II seesaw model, which counts both doubly charged states of that model, and $900$~GeV for the Zee--Babu model, whose single $SU(2)_L$ singlet $k^{\pm\pm}$ has a smaller Drell--Yan cross section than the triplet state~\cite{ATLAS2023Dilepton}.  The PDG lists $1020$~GeV from that same search for the $T_3=\pm1$ state alone~\cite{PDG2026}.  All three assume prompt decays and equal branching fractions into the six same-sign dilepton flavor combinations.  The earlier 36.1~fb$^{-1}$ ATLAS search gives observed $95\%$ confidence-level (CL) limits as a function of the total branching fraction into light leptons~\cite{ATLAS2018Dilepton}, which we translate with the NuFIT~6.0 oscillation ranges~\cite{NuFIT2024} into a flavor-dependent boundary that excludes only where the decays are prompt.  The ATLAS same-sign $WW$ search sets two limits under different assumptions about the charged scalar splitting: $350$~GeV for pair production with the singly charged scalar at least $100$~GeV heavier than the doubly charged one, and $230$~GeV for associated production with that splitting below $5$~GeV, both at unit diboson branching fraction~\cite{ATLAS2021Charged}.  The pair production process itself does not depend on that splitting, so the pair production limit applies wherever the diboson branching fraction is close to one.  That fraction is not close to one everywhere in the survey, so we do not impose the limit as a general exclusion and check it point by point instead.

Charged scalar loops also modify the $h\to\gamma\gamma$ rate~\cite{ArhribDiphoton2012,Chun2012,Dev2013,Ashanujjaman2026}.  We evaluate the leading-order (LO) diphoton signal strength $\mu_{\gamma\gamma}^{\rm LO}$, the ratio of $\sigma(gg\to h)\,{\rm BR}(h\to\gamma\gamma)$, the gluon fusion cross section times the diphoton branching fraction, to its SM value, computed with the physical charged scalar masses and mixing angles.  We require  $0.94\leq\mu_{\gamma\gamma}^{\rm LO}\leq1.24$, the overlap of the separate approximate $2\sigma$ intervals inferred from the ATLAS and CMS measurements~\cite{ATLAS2023Diphoton,CMS2021Diphoton}.  This simple intersection is a screening condition, not a statistical combination of the two measurements.  All the conditions of this section use the tree-level scalar masses of Sec.~\ref{subsec:spectrum}.

\section{One-Loop Effective Potential and Renormalization Group Improvement}
\label{sec:one-loop-rg}

The finite-temperature effective potential considered in this work is given by
\begin{equation}
 V_{\rm eff}=V_0+V_{\rm CW}+V_{\rm ct}+V_{\rm th}+V_{\rm ring}
 ~.
\end{equation}
Here $V_0$ is the tree-level potential of Eq.~\eqref{eq:potential}, while $V_{\rm CW}$, $V_{\rm ct}$, $V_{\rm th}$, and $V_{\rm ring}$ denote the Coleman--Weinberg, finite counterterm, thermal, and ring contributions, respectively.  Every result in this work is evaluated in the \emph{Landau gauge}, where the gauge-fixing parameter $\xi_{\rm gf}=0$ in the $R_\xi$ gauge.  This convention simplifies the gauge--Goldstone sector. A loop-truncated effective potential is still gauge-dependent, and a field coordinate such as the discontinuity $\Delta\varphi/T_c$ quoted below is thus a diagnostic in this gauge, not an observable~\cite{Nielsen1975,Patel2011}.

Bosonic thermal zero modes produce the ring or daisy diagram enhancement.  We include its leading contribution with Arnold--Espinosa resummation.\footnote{In dimensional reduction, the corresponding thermal masses would instead enter the matched three-dimensional theory.}

\subsection{Coleman--Weinberg potential and finite counterterms}
\label{subsec:cw-counterterms}

The field-dependent scalar squared masses are the eigenvalues of the Hessian of the tree-level potential.  We write the entries before imposing the tadpole conditions, so they contain $m^2$, $M^2$ and $\mu$ in place of $M_\Delta^2$.  On a neutral background $(\varphi,\delta)$, the $CP$-even block is
\begin{equation}
 \begin{pmatrix}
 3\lambda_1\varphi^2+\frac12\lambda_{45}\delta^2-m^2-\sqrt2\mu\delta &
 \varphi(\lambda_{45}\delta-\sqrt2\mu)\\
 \varphi(\lambda_{45}\delta-\sqrt2\mu)&
 M^2+3\lambda_{23}\delta^2+\frac12\lambda_{45}\varphi^2
 \end{pmatrix}
 ~.
\end{equation}
The neutral $CP$-odd, singly charged, and doubly charged blocks are treated in the same field basis.  Together with the gauge boson and top quark masses, they enter the Coleman--Weinberg potential
\begin{equation}
 V_{\rm CW}=\sum_i\frac{n_i m_i^4(\varphi,\delta)}{64\pi^2}
 \left[\log\frac{m_i^2(\varphi,\delta)}{Q^2}-c_i\right]
 ~,
\end{equation}
where $n_i$ is the number of real degrees of freedom of species $i$, taken negative for fermions, and $c_i=3/2$ for scalars and fermions and $c_i=5/6$ for gauge bosons in the $\overline{\mathrm{MS}}$ scheme~\cite{ColemanWeinberg1973,Martin2002}.  The real part is used when a background mass is tachyonic.  The omitted imaginary part signals the instability of that background~\cite{WeinbergWu1987}.

Goldstone logarithms make derivatives of the zero-temperature Coleman--Weinberg potential singular as the field-dependent Goldstone mass squared $m_G^2$ tends to zero~\cite{Martin2014}.  We regulate them only in the counterterm conditions and do not resum the Goldstone contributions.  Writing the one-loop kernel of a Goldstone mode as $f(x)=x^2[\log(|x|/Q^2)-3/2]/(64\pi^2)$ with $x=m_G^2$, we differentiate the field-dependent mass first, take the vacuum limit afterward, and impose the counterterm conditions at $Q=Q_{\rm ref}$ with
\begin{equation}
 f'(0)=0
 ~, \qquad
 f''(0)\longrightarrow\frac{1}{32\pi^2}\log\frac{m_h^2}{Q_{\rm ref}^2}
 ~.
 \label{eq:goldstone-log-prescription}
\end{equation}
Here $m_h^2$ is a fixed electroweak reference scale, and any other positive choice defines a different finite prescription.  The infrared problem is established in Ref.~\cite{Martin2014}.  The replacement is our choice and enters only the finite counterterms.

Finite counterterms are written as
\begin{equation}
 V_{\rm ct}=-\frac{\delta m^2}{2}\varphi^2
 +\frac{\delta M^2}{2}\delta^2+\frac{\delta\lambda_1}{4}\varphi^4
 +\frac{\delta\lambda_{23}}{4}\delta^4
 +\frac{\delta\lambda_{45}}{4}\varphi^2\delta^2
 ~.
 \label{eq:counterterm-potential}
\end{equation}
If two tadpole conditions and all three independent neutral curvatures are imposed, the resulting counterterms become singular as $v_\Delta\to0$.  With $B=\partial_\delta V_{\rm CW}$ and $E=\partial_\delta^2V_{\rm CW}$ at the vacuum, those five conditions imply
\begin{equation}
 \delta\lambda_{23}=\frac{B-v_\Delta E}{2v_\Delta^3}
 ~.
 \label{eq:singularct}
\end{equation}
How fast it grows depends on how the limit is taken.  At fixed $M_\Delta$, the trilinear coupling scales as $\mu\propto v_\Delta$.  Unless $B-v_\Delta E$ vanishes as fast as $v_\Delta^3$, the counterterm then grows without bound as $v_\Delta\to0$, and it exceeds the tree-level coupling once $|B-v_\Delta E|>2|\lambda_{23}|v_\Delta^3$.  A thermal history computed with such a large counterterm would be set by the renormalization prescription, outside the control of the loop expansion.

We therefore impose $\delta\lambda_{23}=0$ together with four nonsingular conditions: that the one-loop potential be stationary at the specified vacuum $(v_\Phi,v_\Delta)$, and that the doublet diagonal and doublet--triplet off-diagonal entries of the neutral $CP$-even Hessian take their tree-level values.  These determine $\delta m^2$, $\delta M^2$, $\delta\lambda_1$, and $\delta\lambda_{45}$, and the remaining triplet diagonal Hessian entry becomes a one-loop prediction.  The convention preserves the specified vacuum without putting an inverse power of $v_\Delta$ into a quartic coupling.

\subsection{Renormalization group evolution}
\label{subsec:rg-evolution}

We evolve the couplings from the reference scale $Q_{\rm ref}=246.2$~GeV at one loop, writing $\beta_X\equiv dX/d\log Q$ for every coupling $X$.  The expressions of Refs.~\cite{ChaoZhang2007,Schmidt2007,Chun2012,Dev2013,Bonilla2015,Haba2016,GhoshHighScale2018} use different scalar potential and hypercharge normalizations, and Ref.~\cite{Bonilla2015} identifies a gauge term coefficient that differs from that of Ref.~\cite{Chun2012}.  We translate all of them to the operator basis of Eq.~\eqref{eq:potential} and to $Q_{\rm em}=T_3+Y/2$.  The gauge couplings then obey
\begin{equation}
 16\pi^2\beta_{g_2}=-\frac{5}{2}g_2^3
 ~,\qquad
 16\pi^2\beta_{g'}=\frac{47}{6}g'^3
 ~,\qquad
 16\pi^2\beta_{g_3}=-7g_3^3
 ~.
 \label{eq:gauge-beta-functions}
\end{equation}
The scalar anomalous dimensions are
\begin{equation}
 16\pi^2\gamma_\Phi=3y_t^2-\frac{9}{4}g_2^2-\frac{3}{4}g'^2
 ~, \qquad
 16\pi^2\gamma_\Delta=-6g_2^2-3g'^2
 ~.
 \label{eq:scalar-anomalous-dimensions}
\end{equation}
The top Yukawa coupling obeys
\begin{equation}
 16\pi^2\beta_{y_t}=y_t\left(\frac92y_t^2-\frac{17}{12}g'^2
 -\frac94g_2^2-8g_3^2\right)
 ~.
\end{equation}
In these conventions, the quartic coupling beta functions are
\begin{align}
\begin{split}
16\pi^2\beta_{\lambda_1}={}&
 \frac38g'^4+\frac34g'^2g_2^2+\frac98g_2^4
 -(3g'^2+9g_2^2)\lambda_1+24\lambda_1^2
 +3\lambda_4^2+3\lambda_4\lambda_5
 \\ &\quad
 +\frac54\lambda_5^2+12\lambda_1y_t^2-6y_t^4
 ~,\\
16\pi^2\beta_{\lambda_2}={}&
 6g'^4-12g'^2g_2^2+15g_2^4
 -(12g'^2+24g_2^2)\lambda_2+28\lambda_2^2 +24\lambda_2\lambda_3
 \\ &\quad
 +6\lambda_3^2
 +2\lambda_4^2+2\lambda_4\lambda_5
 ~,\\
 16\pi^2\beta_{\lambda_3}={}&
 24g'^2g_2^2-6g_2^4-(12g'^2+24g_2^2)\lambda_3
 +24\lambda_2\lambda_3+18\lambda_3^2+\lambda_5^2
 ~,\\
 16\pi^2\beta_{\lambda_4}={}&
 3g'^4-6g'^2g_2^2+6g_2^4
 -\left(\frac{15}{2}g'^2+\frac{33}{2}g_2^2\right)\lambda_4
 +12\lambda_1\lambda_4+4\lambda_1\lambda_5
 \\
 &\quad
 +16\lambda_2\lambda_4+12\lambda_3\lambda_4+4\lambda_4^2
 +6\lambda_2\lambda_5 +2\lambda_3\lambda_5+\lambda_5^2+6\lambda_4y_t^2
~,\\
16\pi^2\beta_{\lambda_5}={}&
12g'^2g_2^2
 -\left(\frac{15}{2}g'^2+\frac{33}{2}g_2^2\right)\lambda_5
 +4\lambda_5^2+4\lambda_1\lambda_5+4\lambda_2\lambda_5
 \\
 &\quad
 +8\lambda_3\lambda_5
 +8\lambda_4\lambda_5+6\lambda_5y_t^2
 ~.
 \end{split}
 \label{eq:quartic-beta-functions}
\end{align}
We add a field-independent term $\Omega(Q)$ as the vacuum energy to $V_0$, and its running is required for the RG invariance of the complete potential~\cite{Ford1993}.  At vanishing fields, the scalar squared masses are $-m^2$ for the four doublet components and $M^2$ for the six triplet components, so the one-loop potential changes with the scale as $dV_{\rm CW}/d\log Q=-(4m^4+6M^4)/(32\pi^2)$, which $\beta_\Omega$ below cancels.  We fix $\Omega(Q_{\rm ref})=0$.  A field-independent term would drop out of every difference of the potential at a fixed scale, but the field-dependent scale of Eq.~\eqref{eq:central-running-scale} makes $\Omega[Q(\varphi,\delta,T)]$ enter the potential differences and derivatives.  The mass and trilinear parameters and $\Omega$ obey
\begin{align}
\begin{split}
 16\pi^2\beta_{m^2}={}&m^2\left(12\lambda_1+6y_t^2
 -\frac{9}{2}g_2^2-\frac{3}{2}g'^2\right)
 -3M^2(2\lambda_4+\lambda_5)-12\mu^2
 ~,\\
 16\pi^2\beta_{M^2}={}&M^2\left(16\lambda_2+12\lambda_3
 -12g_2^2-6g'^2\right)
 -2m^2(2\lambda_4+\lambda_5)+4\mu^2
 ~,\\
 16\pi^2\beta_\mu={}&\mu\left(4\lambda_1+4\lambda_4+6\lambda_5
 +6y_t^2-\frac{21}{2}g_2^2-\frac{9}{2}g'^2\right)
 ~,\\
 16\pi^2\beta_\Omega={}&2m^4+3M^4
 ~.
 \end{split}
 \label{eq:dimensionful-beta-functions}
\end{align}

The relation $\beta_\mu\propto\mu$ follows from the restoration of lepton number symmetry in the limit $\mu\to0$, and the running cannot generate a nonzero $\mu$ from $\mu=0$.  At fixed scalar masses and quartic couplings, Eq.~\eqref{eq:MDelta-distinction} gives $v_\Delta\propto\mu$.  A small triplet VEV is thus technically natural in the sense of 't Hooft~\cite{tHooft1980}.  We implement RG improvement using the standard effective potential organization in which running parameters and field factors resum logarithms fixed by the Callan--Symanzik equation~\cite{Ford1993}.

The renormalization scale of the RG-improved calculation is the function
\begin{equation}
 Q(\varphi,\delta,T)=\sqrt{\varphi^2+2\delta^2+(\pi T)^2}
 ~.
 \label{eq:central-running-scale}
\end{equation}
Here, $\varphi^2+2\delta^2$ is the squared neutral radial norm in the field normalization of Eq.~\eqref{eq:neutralpotential}, the combination that gives the $W$ boson mass $m_W^2=g_2^2(\varphi^2+2\delta^2)/4$, and $\pi T$ is the lowest fermionic Matsubara frequency.  A Matsubara frequency is one of the discrete frequencies allowed by periodic or antiperiodic fields in imaginary thermal time.  The scale $Q$ follows the largest background or thermal scale and varies continuously between the potentials at zero and finite temperature.  The couplings are evolved only over $60~\mathrm{GeV}\leq Q_{\rm ren}\leq5~\mathrm{TeV}$, with $Q_{\rm ren}$ the renormalization scale defined below.  Below the lower limit, both the running parameters and the scale of the explicit loop logarithms are held at their values at $60$~GeV, because the truncated one-loop evolution would otherwise amplify infrared logarithms.  Phases that approach the upper limit are excluded from the analysis.  Both limits are part of the prescription.  To vary the scale, we use $Q_{\rm ren}=\kappa Q(\varphi,\delta,T)$ with $0.50\leq\kappa\leq1.50$, both with and without RG evolution to study the RG improvement.  Without the improvement, the Lagrangian parameters keep their values at $Q_{\rm ref}$, the field renormalization factors are not evolved, and only $Q_{\rm ren}$ in the explicit loop logarithms varies.  We refer to the choice $\kappa=1$ in Landau gauge with Arnold--Espinosa resummation and RG improvement as the \emph{central prescription}.  The finite-order zero-temperature minimum can move as $\kappa$ varies, so the scale variation is not an uncertainty band at a fixed physical vacuum.  In both treatments, the finite counterterms are those fixed at $Q_{\rm ref}$, and at $\kappa=1$ and $T=0$ the logarithms are evaluated at $Q(v)=Q_{\rm ref}$.  Thus, both reproduce the electroweak vacuum there, and the comparison isolates the running of the parameters.

\subsubsection{One-loop power counting}
\label{subsubsec:power-counting}

The running fields are $\phi_i(Q)=Z_i(Q)\phi_i$, with $Z_i(Q_{\rm ref})=1$ and
\begin{equation}
 \frac{d\log Z_i}{d\log Q}=-\gamma_i
 ~, \qquad
 i=\Phi,\Delta
 ~.
 \label{eq:field-running}
\end{equation}
Write the effective potential as $V=V_0+\hbar(V_1+V_{\rm ct})$, with $V_1$ collecting the Coleman--Weinberg, thermal, and ring terms, $V_{\rm ct}$ the counterterm potential of Eq.~\eqref{eq:counterterm-potential}, and $\hbar$ counting loops.  Expanding a running coupling and a running field around $Q_{\rm ref}$ gives
\begin{equation}
 X(Q)=X+\hbar\beta_{X} L+\mathcal{O}(\hbar^2)
 ~, \qquad
 Z_i(Q)=1-\hbar\gamma_i L+\mathcal{O}(\hbar^2)
 ~, \qquad
 L=\log\frac{Q}{Q_{\rm ref}}
 ~.
\end{equation}
A change in the arguments of the one-loop contribution first enters at
\begin{equation}
 \hbar^2 L\left(\beta_{X}\frac{\partial}{\partial X}
 -\sum_i\gamma_i\phi_i\frac{\partial}{\partial\phi_i}\right)V_1
 ~,
 \label{eq:rgi-higher-order-shift}
\end{equation}
which lies beyond one loop.  We consequently evaluate the tree-level potential with running couplings and field factors.  The Coleman--Weinberg, thermal, and ring terms also use the running Lagrangian parameters at $Q$, but they are evaluated on the classical fields.  So is the counterterm potential, whose coefficients stay fixed at $Q_{\rm ref}$.

The field-dependent scale produces a formally higher-order radial tadpole at the specified vacuum.  With $Q^2=\varphi^2+2\delta^2$ at $T=0$, the gradient of $\log Q$ is proportional to $(\varphi,2\delta)$, so the induced tadpole at the vacuum points along $(v_\Phi,2v_\Delta)$.  We remove it by adding to the counterterms
\begin{equation}
 \Delta V_{\rm rem}=-\frac{\Delta m_r^2}{2}\varphi^2
 +\frac{\Delta M_r^2}{2}\delta^2
 ~, \quad
 \Delta m_r^2=\left.\frac{1}{v_\Phi}
 \frac{\partial V_{\rm strict}}{\partial\varphi}\right|_{v_\Phi,v_\Delta,T=0}
 ~, \quad
 \Delta M_r^2=-2\Delta m_r^2
 ~,
\label{eq:radial-rematching}
\end{equation}
where $V_{\rm strict}$ denotes the zero-temperature potential evaluated with the field-dependent scale before $\Delta V_{\rm rem}$ is added.  Therefore, $\Delta V_{\rm rem}=-\Delta m_r^2(\varphi^2+2\delta^2)/2$ is quadratic in the same neutral radial norm that sets the scale.  The two coefficients are calculated once for each parameter point with the central-scale rule and then held fixed as the temperature or $\kappa$ changes.  This minimal rematching cancels the radial tadpole at $T=0$ but does not refit $m_h$ or the complete two-dimensional curvature matrix.  The potential without RG improvement carries the same term, which removes a tadpole of one-loop order there.

\section{Finite-Temperature Effective Potential}
\label{sec:thermal-potential}
\subsection{Thermal functions}
\label{subsec:thermal-functions}

For $y=m_i^2/T^2\geq0$, with $m_i$ a field-dependent mass, the standard one-loop thermal functions are~\cite{DolanJackiw1974}
\begin{align}
\begin{split}
 J_B(y)&=\int_0^\infty dx\,x^2\log\!\left(1-e^{-\sqrt{x^2+y}}\right)
 ~,\\
 J_F(y)&=\int_0^\infty dx\,x^2\log\!\left(1+e^{-\sqrt{x^2+y}}\right)
 ~.
\end{split}
\end{align}
The thermal contribution to the potential is $V_{\rm th}=(T^4/2\pi^2)\sum_in_iJ_{B,F}\!\left(m_i^2(\varphi,\delta)/T^2\right)$, with $J_B$ for bosons and $J_F$ for fermions and the same counting of degrees of freedom $n_i$ used in the Coleman--Weinberg potential.  Expanding the logarithms and integrating term by term gives the Bessel series representation
\begin{align}
\begin{split}
 J_B(y)&=-\sum_{n=1}^{\infty}\frac{y}{n^2}K_2(n\sqrt y)
 ~,\\
 J_F(y)&=\sum_{n=1}^{\infty}(-1)^{n+1}\frac{y}{n^2}K_2(n\sqrt y)
 ~,
\end{split}
\end{align}
where $K_2$ is the modified Bessel function of the second kind of order two.  Along an unstable background direction, a scalar mass squared can be negative.  For $y<0$, we use the real part of the continued thermal function~\cite{Wainwright2012}, held at its value at its first minimum, $y\simeq-3.7$, for more negative arguments.  At the completed points of Sec.~\ref{sec:completed-transition-benchmarks}, no squared mass that enters $J_B$ between their nucleation and critical temperatures is negative, so neither choice affects them.  The imaginary part is not included in the minimization.  The integral and series representations reproduce the analytic limits
\begin{equation}
 J_B(0)=-\frac{\pi^4}{45}
 ~,\qquad
 J_F(0)=\frac{7\pi^4}{360}
 ~,
\end{equation}
and, for $y\to0^+$, the high-temperature expansions
\begin{align}
\begin{split}
 J_B(y)={}&-\frac{\pi^4}{45}+\frac{\pi^2}{12}y
 -\frac{\pi}{6}y^{3/2}-\frac{y^2}{32}\log\frac{y}{a_b}
 +\mathcal{O}(y^{5/2})
 ~,\\
 J_F(y)={}&\frac{7\pi^4}{360}-\frac{\pi^2}{24}y
 -\frac{y^2}{32}\log\frac{y}{a_f}+\mathcal{O}(y^3)
 ~,
\end{split}
 \label{eq:thermal-high-temperature-expansions}
\end{align}
where
\begin{equation}
 a_b=16\pi^2\exp\!\left(\frac32-2\gamma_E\right)
 ~,\qquad
 a_f=\pi^2\exp\!\left(\frac32-2\gamma_E\right)
 ~,
\end{equation}
and $\gamma_E$ is the Euler--Mascheroni constant.  Only the bosonic expansion contains a cubic term in the mass that can produce a thermal barrier, while ring resummation screens the corresponding zero mode.  The numerical potential uses the full thermal functions.

\subsection{Leading thermal self-energies}
\label{subsec:self-energies}

For a canonically normalized real scalar $\phi_a$, the scalar loop contribution is obtained directly from the quartic tensor of $V_4$, the quartic part of the potential of Eq.~\eqref{eq:potential} written in the ten real scalar components,
\begin{equation}
 \frac{\Pi^{\rm scalar}_{aa}}{T^2}
 =\frac{1}{24}\sum_c
 \left.\frac{\partial^4 V_4}
 {\partial\phi_a^2\partial\phi_c^2}\right|_{\phi=0}
 ~.
\end{equation}
Carrying out the sum gives the scalar parts $\Pi_\Phi^{\rm scalar}=T^2(8\lambda_1+4\lambda_4+2\lambda_5)/16$ and $\Pi_\Delta^{\rm scalar}=T^2(8\lambda_2+6\lambda_3+2\lambda_4+\lambda_5)/12$, while the gauge and top Yukawa contributions are $\Pi_\Phi^{\rm gauge+Yukawa}=T^2(3g_2^2+g'^2+4y_t^2)/16$ and $\Pi_\Delta^{\rm gauge}=T^2(6g_2^2+3g'^2)/12$.  Adding them gives
\begin{align}
\begin{split}
 \Pi_\Phi&=\frac{T^2}{16}
 \left[3g_2^2+g'^2+8\lambda_1+4\lambda_4+2\lambda_5+4y_t^2\right]
 ~,\\
 \Pi_\Delta&=\frac{T^2}{12}
 \left[6g_2^2+3g'^2+8\lambda_2+6\lambda_3+2\lambda_4+\lambda_5\right]
 ~.
\end{split}
\end{align}
Every real component of a given multiplet has the same coefficient, as required by the unbroken gauge symmetry at the origin.

Each longitudinal coefficient is a weighted count of the thermal matter content, $\Pi_{W_L}/(g_2^2T^2)=\left[2C_A+\sum_fT(R_f)+2\sum_sT(R_s)\right]/6$ and $\Pi_{B_L}/(g'^2T^2)=\left[\tfrac14\sum_fY_f^2+\tfrac12\sum_sY_s^2\right]/6$, where $C_A=2$ for $SU(2)_L$, $T(R)$ is the Dynkin index, the fermion sums run over the Weyl fermions and the scalar sums over the complex scalars.  Including the full thermal matter content gives
\begin{equation}
 \Pi_{W_L}=\frac52 g_2^2T^2
 ~,\qquad
 \Pi_{B_L}=\frac{17}{6}g'^2T^2
 ~.
\end{equation}
Because the two coefficients differ, the Debye masses are not diagonal in the zero-temperature $\gamma$--$Z$ basis, so they must be added to the neutral longitudinal mass matrix before it is diagonalized (Sec.~\ref{subsec:neutral-screening}).

\subsection{Range of the high-temperature approximation}
\label{subsec:high-t-range}

\begin{figure}[htbp]
\centering
\includegraphics[width=0.8\textwidth]{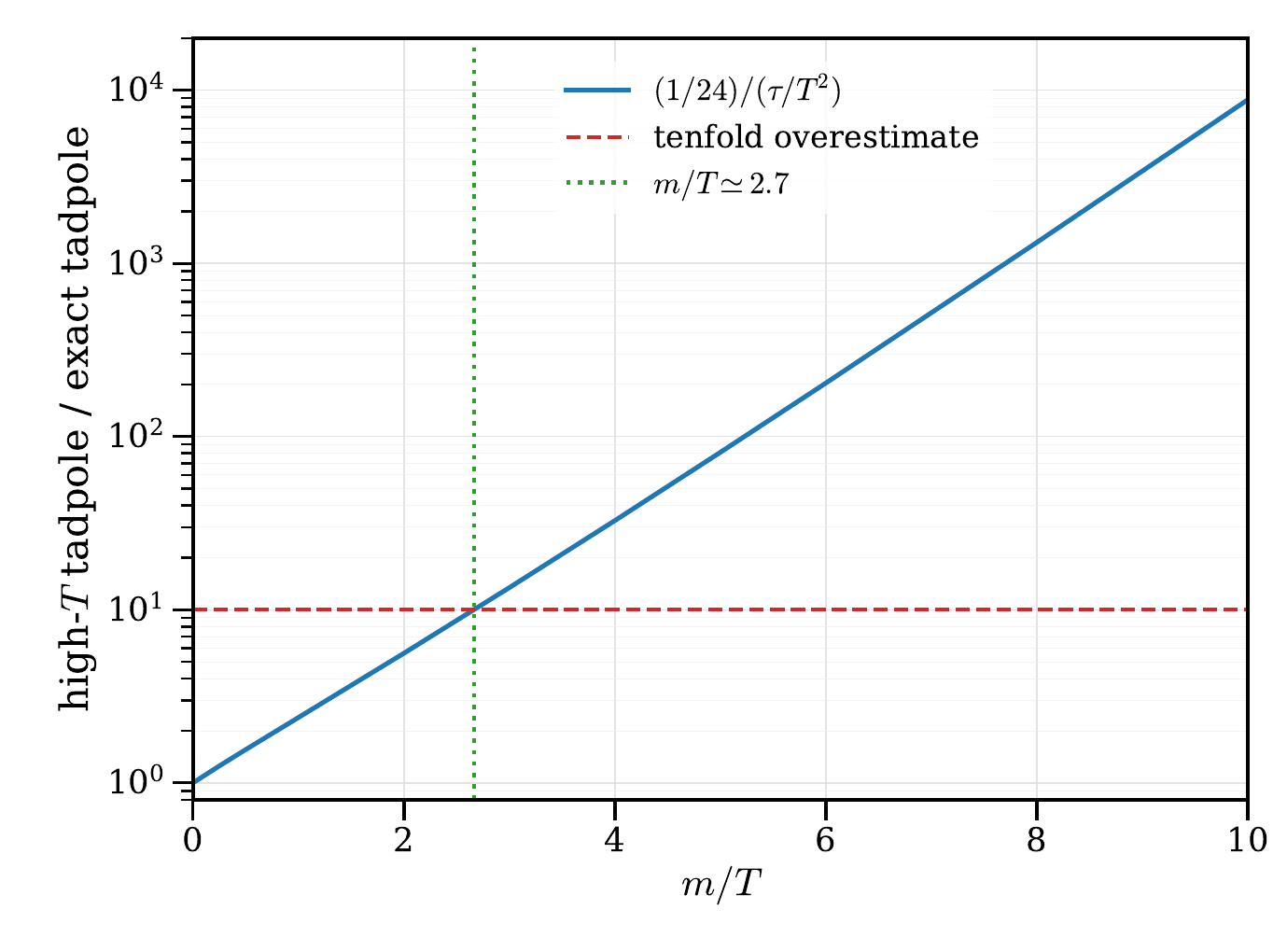}
\caption{Ratio of the massless high-temperature tadpole to the exact massive one-loop result as a function of $m/T$.  The blue line joins direct numerical evaluations of the integral above.}
\label{fig:tadpole}
\end{figure}

The additional mostly triplet scalar modes need not be light compared with the plasma temperature, whereas their Debye coefficients are obtained in the massless high-temperature limit.  The exact one-loop bosonic tadpole per real degree of freedom, $\tau(m,T)\equiv\partial[T^4J_B(m^2/T^2)/(2\pi^2)]/\partial m^2$, is
\begin{equation}
 \frac{\tau(m,T)}{T^2}=\frac{1}{4\pi^2}\int_0^\infty
 \frac{x^2\,dx}{\sqrt{x^2+z^2}\,[e^{\sqrt{x^2+z^2}}-1]}
 =\frac{z}{4\pi^2}\sum_{n=1}^{\infty}\frac{K_1(nz)}{n}
 ~,
 \qquad z=\frac{m}{T}
 ~,
\end{equation}
with $K_1$ the modified Bessel function of the second kind of order one.  The Debye coefficients above use the massless limit $\tau/T^2=1/24$.  The massless limit overestimates the exact tadpole by a factor of ten at $m/T\simeq2.7$ and by nearly four orders of magnitude at $m/T=10$ (Fig.~\ref{fig:tadpole}).

A state heavier than the plasma temperature therefore contributes to the Debye masses of the light modes as though it were much lighter, although its thermal population is Boltzmann suppressed.  A single bosonic tadpole sets no universal cutoff in $m/T$ for a potential whose species differ in mass, multiplicity, and field dependence.  It does show that the massless screening coefficient cannot be assumed to track thermal decoupling uniformly over the mass range of the mostly triplet states.  Since the same Debye coefficients set the zero-mode masses in the ring term, missed decoupling can alter the barrier between phases.  We use the exact tadpole only as a diagnostic.  It contains both hard and soft contributions, and inserting it in the zero-mode ring term would mix them.  A consistent resummation instead matches the hard contributions into the soft theory and subtracts the overlap order by order~\cite{Croon2021,Ekstedt2023}.

\subsection{Neutral screening and the ring replacement}
\label{subsec:neutral-screening}

On the neutral backgrounds, the zero-temperature gauge matrix is
\begin{equation}
 \mathcal{M}_0^2
 =
 \left(\frac{\varphi^2}{4}+\delta^2\right)
 \begin{pmatrix}g_2^2&-g_2g'\\-g_2g'&g'^2\end{pmatrix}
 ~.
\end{equation}
The longitudinal matrix is
\begin{equation}
 \mathcal{M}_L^2
 =
 \mathcal{M}_0^2+
 \begin{pmatrix}\Pi_{W_L}&0\\0&\Pi_{B_L}\end{pmatrix}
 ~,
\end{equation}
and must be diagonalized only after this addition.  Let $s_W=g'/\sqrt{g_2^2+g'^2}$ and $c_W=g_2/\sqrt{g_2^2+g'^2}$, and define the $\gamma$ and $Z$ fields by $A_\mu=s_WW^3_\mu+c_WB_\mu$ and $Z_\mu=c_WW^3_\mu-s_WB_\mu$.  Rotating the Debye term to the zero-temperature $\gamma$--$Z$ basis gives
\begin{equation}
 \Pi_{AZ}=s_Wc_W(\Pi_{W_L}-\Pi_{B_L})
 =\frac{g_2g'}{g_2^2+g'^2}
 (\Pi_{W_L}-\Pi_{B_L})\neq0
 ~.
\end{equation}
At the symmetric origin, the eigenvalues are $\Pi_{W_L}$ and $\Pi_{B_L}$, not a massless longitudinal photon and a single screened $Z$.

For each bosonic zero mode $i$, with $n_i$ and $m_i^2(\varphi,\delta)$ as in the Coleman--Weinberg potential of Sec.~\ref{subsec:cw-counterterms} and $\Pi_i(T)$ its thermal self-energy, the Arnold--Espinosa prescription adds~\cite{ArnoldEspinosa1993}
\begin{equation}
 V_{\rm ring}^{(i)}=-\frac{n_iT}{12\pi}
 {\left\{\operatorname{Re}\!\left[(m_i^2+\Pi_i)^{3/2}\right]
 -\operatorname{Re}\!\left[(m_i^2)^{3/2}\right]\right\}}
 ~.
\end{equation}
Because a background mass squared can be negative, the fractional powers are evaluated on the principal branch before their real parts are taken.  The second term cancels the unresummed $-n_iTm_i^3/(12\pi)$ already contained in the thermal function, so the cubic term enters once, in its screened form.

The ring sum runs over the ten real scalar modes.  Because the doublet and triplet components mix on a background with $\delta\neq0$, $\Pi_\Phi$ is added to the doublet and $\Pi_\Delta$ to the triplet diagonal entries of the scalar mass matrix before it is diagonalized.  The resummed term uses the eigenvalues of this screened matrix.  The ring sum also includes the two charged longitudinal gauge modes with $\Pi_i=\Pi_{W_L}$.  For the two neutral longitudinal modes, the resummed term uses the eigenvalues of $\mathcal{M}_L^2$ and the subtracted unresummed term uses the eigenvalues of $\mathcal{M}_0^2$.  Transverse gauge modes are not included in the ring replacement, since their zero-mode self-energy has no term of order $g_2^2T^2$ and the magnetic mass of order $g_2^2T$ that screens them in the non-Abelian theory is not calculable in perturbation theory.

\section{Thermal Histories, Tunneling Actions, and Cosmological Criteria}
\label{sec:ct-methods}

\subsection{Thermal histories}
\label{subsec:thermal-histories}

We use CT version~2.0.7~\cite{Wainwright2012,CosmoTransitions207} to trace thermal minima, construct the phase structure, locate critical transitions, and for comparison compute two-field bounce actions by path deformation.  The actions entering the nucleation integrals are approximated by shooting along the straight path between the two minima and checked against the full two-field path deformation of CT on the same minima.  All of this is done on the neutral backgrounds $(\varphi,\delta)$ of Eq.~\eqref{eq:neutralpotential}, with every charged field set to zero.  Every phase, tunneling solution and statement about a global minimum in this paper refers to that two-dimensional field space.  Thermal phases that break electric charge lie outside the present calculation.  The three-dimensional bounce action $S_3(T)$ is the $O(3)$-symmetric saddle point action of the two-field bubble profile, and has dimensions of energy so that $S_3/T$ is dimensionless.  The straight-path result is its restricted-path approximation.  A single parameter point may contain several phases and transitions.  We call the phase occupied by the Universe immediately before a candidate transition the \emph{parent phase} and the one immediately after the \emph{daughter phase}.  A pairwise degeneracy corresponds to a completed FOPT only if the cooling history reaches it through the parent phase and bubbles of the competing phase subsequently nucleate, percolate, and complete the transition.

\subsection{\texorpdfstring{Nucleation, percolation, and completion}{Nucleation, percolation, and completion}}
\label{subsec:nucleation-completion-gw}

The frequently used condition $S_3/T=140$ is motivated near the electroweak scale by $\Gamma/H^4\sim1$, with $\Gamma$ the nucleation rate per unit volume and $H$ the Hubble rate, approximately one nucleated bubble per Hubble volume per Hubble time.  It presupposes both the temperature scale and the expansion rate.  We use it only as a local diagnostic, and fix the nucleation temperature by integrating the nucleation rate instead.

The semiclassical bounce formalism originates with the zero-temperature false-vacuum decay calculation of Refs.~\cite{Coleman1977,CallanColeman1977}, and its thermal $O(3)$ form was established in Ref.~\cite{Linde1983}.  We approximate the nucleation rate per unit volume by~\cite{Linde1983,Caprini2020}
\begin{equation}
 \Gamma(T)
 =
 T^4\left[\frac{S_3(T)}{2\pi T}\right]^{3/2}
 \exp\!\left[-\frac{S_3(T)}{T}\right]
 ~.
 \label{eq:thermal-nucleation-rate}
\end{equation}
For a Universe dominated by radiation, the false-vacuum fraction is given by~\cite{EllisLewickiNo2019,Ellis2020}
\begin{align}
\begin{split}
 P_{\rm false}(T)
 &=
 e^{-I(T)}
 ~, \\
 I(T)
 &=
 \frac{4\pi}{3}\int_T^{T_c}
 \frac{dT'\,\Gamma(T')a^3(T')}{T'H(T')}
 r^3(T,T')
 ~,~
 r(T,T')
 =
 \int_T^{T'}\frac{dT''\,v_w}{T''H(T'')a(T'')}
 ~.
\end{split}
 \label{eq:false-vacuum-fraction}
\end{align}
Here $r(T,T')$ is the comoving radius at temperature $T$ of a bubble nucleated at $T'$, $v_w$ is the bubble wall speed in the plasma frame, $a(T)$ is the scale factor, and $H(T)$ is the Hubble rate.  The Friedmann equation gives $H^2=8\pi\rho_{\rm rad}/(3M_{\rm Pl}^2)$, with the Planck mass $M_{\rm Pl}$ and the radiation energy density $\rho_{\rm rad}=\pi^2g_*T^4/30$, where $g_*$ is a temperature-independent count of relativistic degrees of freedom.  We set $g_*=106.75$, the SM value with every species relativistic.  We do not compute the wall speed.  It is set by the nonequilibrium friction of the plasma on the wall, which an equilibrium potential does not determine~\cite{Espinosa2010,Caprini2020}.  For the completed transitions, we take two representative values $v_w=0.6$ and $0.95$.  We hold the relativistic energy and entropy counts fixed, so $a(T)\propto1/T$ and $dT/dt=-H(T)T$.  The upper limit is $T_c$ because bubbles of the daughter phase can nucleate only where it has the lower free energy.  We define the percolation temperature $T_p$ by the condition $I(T_p)=0.34$ used in Ref.~\cite{EllisLewickiNo2019}, the continuum percolation threshold at which overlapping bubbles first form a cluster that spans the volume.  We also record $I=1$, where $P_{\rm false}=e^{-1}$, the percolation condition adopted in Ref.~\cite{Ellis2020}.  Our completion condition is $I=\log100$, where $P_{\rm false}=0.01$ and 99\% of the volume has converted.  The physical false-vacuum volume is proportional to $a^3P_{\rm false}$ and decreases only when
\begin{equation}
 \frac{d\log(a^3P_{\rm false})}{dt}
 =H\left(3+T\frac{dI}{dT}\right)<0
 ~.
 \label{eq:false-vacuum-volume-decrease}
\end{equation}
We test this condition, which is the completion criterion of Ref.~\cite{EllisLewickiNo2019}, in addition to the converted fraction.

The expected number of bubbles nucleated per Hubble volume is $N(T)=\int_T^{T_{\rm max}}dT'\,\Gamma(T')/[T'H^4(T')]$.  The nucleation temperature is fixed by $N(T_n)=1$.  The $N$ and $I$ integrals are evaluated only up to $T_{\rm max}$, the highest temperature at which a converged action was stored, which lies a fraction of a GeV below $T_c$ at the completed points.  There $S_3/T$ already exceeds $800$ and is still rising toward $T_c$.  If it keeps rising, the interval above $T_{\rm max}$ contributes nothing measurable to $N$ or $I$, and the nucleation, percolation and completion results assume that it does.

A point with no root of the integrated nucleation condition need not stay in the metastable phase.  If the barrier disappears first, the phase becomes spinodally unstable and the field rolls classically.  The vacuum can then change without the bubble population that the acoustic GW formula assumes.  Any completion analysis must therefore keep \emph{nucleated} and \emph{spinodal roll} as distinct outcomes.

\subsection{\texorpdfstring{Strength, inverse duration, and radiation count}{Strength, inverse duration, and radiation count}}
\label{subsec:strength-duration}

With $\Delta V=V_{\rm false}-V_{\rm true}$ and the $\rho_{\rm rad}$ given in the previous subsection, the trace anomaly strength and inverse duration used in the spectra are
\begin{align}
 \alpha_\theta
 &=\left.\frac{\Delta V-(T/4)\,d\Delta V/dT}{\rho_{\rm rad}}
 \right|_{T_*}
 ~~~\mbox{and}~~~
 \frac{\beta}{H_*}
 =T_*\left.\frac{d(S_3/T)}{dT}\right|_{T_*}
 ~.
 \label{eq:gw-alpha-beta}
\end{align}
Here, $T_*=T_n$ for the completed transition analysis, and $H_*=H(T_*)$ is the Hubble rate at that temperature.  The inverse duration is $\beta\equiv-d(S_3/T)/dt$ at $T_*$, which $dT/dt=-HT$ turns into the temperature derivative above.  The combination in $\alpha_\theta$ is the relativistic-plasma limit of the pseudotrace strength used for a general equation of state~\cite{Giese2020}.  The efficiency fit of Ref.~\cite{Espinosa2010} assumes a bag equation of state, with the squared sound speed $c_s^2=1/3$ in each phase, so combining it with $\alpha_\theta$ is an approximation.  Recent analyses recommend $T_p$ instead, as more directly tied to transition progress and bubble collisions~\cite{Athron2023}.  Our use of $T_n$ is a convention, and it is a good approximation at the completed points of Sec.~\ref{sec:completed-transition-benchmarks}.  There $T_p$ lies below $T_n$ by less than $0.1\%$ at both wall speeds, from $0.3$~GeV at $P_1$ to $0.01$~GeV at $P_7$.

For the energy and entropy densities, we use the relativistic SM count $g_*=106.75$ throughout.  Counting the six real triplet degrees of freedom as relativistic would give $112.75$, which at fixed temperature lowers $\alpha_\theta$ by $5\%$ and raises $H$ by $3\%$.  We have not propagated this into $T_n$ and $T_p$.

\subsection{\texorpdfstring{GW sources and spectra}{GW sources and spectra}}
\label{subsec:gw-spectra}

We report the sound wave contribution as the primary GW prediction.  A magnetohydrodynamic turbulence contribution is shown separately under the assumption $\epsilon_{\rm turb}=\kappa_{\rm turb}/\kappa_{\rm sw}=0.10$ for illustration, where $\kappa_{\rm sw}$ and $\kappa_{\rm turb}$ are the fractions of the released energy that end up in bulk fluid motion and in turbulent motion.  The scalar-field collision contribution matters when a runaway wall retains an appreciable fraction of the released energy.  A wall that is driven by the plasma and does not run away reaches a terminal speed and transfers most of that energy to the fluid~\cite{BodekerMoore2009,BodekerMoore2017,Caprini2020}.  The collision amplitude carries the factor $[\kappa_\phi\alpha_\theta/(1+\alpha_\theta)]^2$, with $\kappa_\phi$ the fraction of the released energy retained in the scalar field.  Even at $\kappa_\phi=1$ this factor is no larger than $\alpha_\theta^2$, and the completed transitions found here have $\alpha_\theta$ of order $10^{-5}$.  Small $\alpha_\theta$ alone does not exclude a runaway wall.  That depends on the microscopic plasma friction and the driving pressure, neither of which follows from minimizing an equilibrium potential, and we have not computed them point by point.  Omitting the collision term is therefore an assumption that the walls do not run away, combined with the parametric bound above.

We write $\widetilde\beta=\beta/H_*$ and $S_i=S_i(f/f_i)$, with $f$ the frequency observed today and $f_i$ the characteristic frequency of component $i$, and $v_{\rm sh}\equiv\max(v_w,c_s)$, with $c_s=1/\sqrt3$ the sound speed of a relativistic plasma.  For the sound waves we use the fit to acoustic simulations~\cite{Hindmarsh2014,Hindmarsh2015,Hindmarsh2017}, with the amplitude normalization and the redshifted peak frequency in the form collected in Ref.~\cite{Athron2024Review}.  The spectral shapes are those of Ref.~\cite{Caprini2016}, and the turbulence form is that of Ref.~\cite{CapriniDurrerServant2009} in the same normalization.  The acoustic lifetime prescription paired with that normalization replaces the sharp cutoff of Refs.~\cite{Caprini2020,Ellis2020} by a form that approaches it smoothly.  The two components are
\begin{align}
\begin{split}
 h^2\Omega_{\rm sw}(f)
 &=2.59\times10^{-6}\,\Upsilon
 \frac{v_{\rm sh}}{\widetilde\beta}
 \left(\frac{\kappa_{\rm sw}\alpha_\theta}{1+\alpha_\theta}\right)^2
 \left(\frac{100}{g_*}\right)^{1/3}S_{\rm sw}
 ~, \\
 h^2\Omega_{\rm turb}(f)
 &=3.35\times10^{-4}
 \frac{v_w}{\widetilde\beta}
 \left(\frac{\kappa_{\rm turb}\alpha_\theta}{1+\alpha_\theta}\right)^{3/2}
 \left(\frac{100}{g_*}\right)^{1/3}S_{\rm turb}
 ~.
\end{split}
 \label{eq:gw-components}
\end{align}
Here $h$ is the present Hubble rate in units of $100$~km\,s$^{-1}$\,Mpc$^{-1}$.  The corresponding spectral shapes, in terms of $x=f/f_i$ and the Hubble rate $h_*$ at $T_*$ redshifted to today, given below, are
\begin{align}
 S_{\rm sw}(x)&=x^3\left(\frac{7}{4+3x^2}\right)^{7/2}~,\notag\\
 S_{\rm turb}(x)&=\frac{x^3}{(1+x)^{11/3}
 [1+8\pi f/h_*]}~.
 \label{eq:gw-shapes}
\end{align}
Redshifted from $T_*$ to the present by entropy conservation, with the degrees of freedom evaluated at $T_*$, the characteristic frequencies observed today are
\begin{align}
\begin{split}
 f_{\rm sw}&=8.9\times10^{-6}\,{\rm Hz} \,
 \frac{\widetilde\beta}{v_{\rm sh}}
 \left(\frac{T_*}{100\,{\rm GeV}}\right)
 \left(\frac{g_*}{100}\right)^{1/6}
 ~, \\
 f_{\rm turb}&=2.7\times10^{-5}\,{\rm Hz} \,
 \frac{\widetilde\beta}{v_w}
 \left(\frac{T_*}{100\,{\rm GeV}}\right)
 \left(\frac{g_*}{100}\right)^{1/6}
 ~, \\
 h_*&=1.65\times10^{-5}\,{\rm Hz}\,
 \left(\frac{T_*}{100\,{\rm GeV}}\right)
 \left(\frac{g_*}{100}\right)^{1/6}
 ~.
\end{split}
 \label{eq:gw-frequencies}
\end{align}
The finite lifetime of the acoustic source is included through the suppression factor $\Upsilon$, in which $\tau_{\rm sw}$ is the lifetime of the sound wave source, $R_*$ the mean bubble separation at $T_*$, and $\overline U_f$ the root-mean-square bulk fluid velocity,
\begin{align}
\begin{split}
 \Upsilon
 &=1-\frac{1}{\sqrt{1+2H_*\tau_{\rm sw}}}
 ~, \qquad
 H_*\tau_{\rm sw}=\frac{H_*R_*}{\overline U_f}
 ~, \\
 H_*R_*
 &=
 (8\pi)^{1/3}\frac{v_{\rm sh}}{\widetilde\beta}
 ~, \qquad
 \overline U_f=\sqrt{\frac{3}{4}
 \frac{\kappa_{\rm sw}\alpha_\theta}{1+\alpha_\theta}}
 ~.
\end{split}
 \label{eq:gw-acoustic-lifetime}
\end{align}
For a subsonic wall, the larger of $v_w$ and $1/\sqrt3$ enters because the fluid shell sets the size of the acoustic source~\cite{Caprini2020,Ellis2020}.  The factor $\Upsilon$ is the one derived in Ref.~\cite{Guo2021} for an acoustic source that decays in an expanding Universe.  For a short-lived source, $H_*\tau_{\rm sw}\ll1$, it reduces to $\Upsilon\simeq H_*\tau_{\rm sw}$, which is the sharp cutoff $\min(1,H_*\tau_{\rm sw})$ used in Refs.~\cite{Caprini2020,Ellis2020}.  For $H_*\tau_{\rm sw}\gg1$ it approaches one only as $1-(2H_*\tau_{\rm sw})^{-1/2}$.  The two forms differ most near $H_*\tau_{\rm sw}\sim1$.  The amplitude coefficient $2.59\times10^{-6}$ and the frequency coefficient $8.9\times10^{-6}$~Hz are those of Ref.~\cite{Athron2024Review}, which follow from the simulations of Refs.~\cite{Hindmarsh2015,Hindmarsh2017} with the correction given in Ref.~\cite{Hindmarsh2017}.  The frequency coefficient places the acoustic peak at $k_pR_*=10$, the wavenumber where the simulated spectra peak.  The fluid efficiency $\kappa_{\rm sw}(\alpha_\theta,v_w)$, the fraction of the released energy converted into bulk motion of the plasma, is evaluated with the full deflagration, hybrid, and detonation interpolation of Ref.~\cite{Espinosa2010}.  We then set $\kappa_{\rm turb}=\epsilon_{\rm turb}\kappa_{\rm sw}$ only for the separately labeled turbulence estimate.  Of the two representative wall speeds, $v_w=0.6$ lies just above the ideal radiation fluid sound speed $c_s=1/\sqrt3\simeq0.577$ and gives a moderately fast detonation for these weak transitions, and $v_w=0.95$ is a fast wall near the relativistic limit but not a runaway one.  Neither choice determines $\kappa_\phi$.  A preferred wall speed, or a test for runaway behavior, needs the nonequilibrium plasma friction~\cite{Espinosa2010,Caprini2020}.

\subsection{\texorpdfstring{Sphaleron washout criterion}{Sphaleron washout criterion}}
\label{subsec:sphaleron-criterion}

The commonly quoted measure of the strength of a transition is the ratio of the field value to the temperature, $\xi_T\equiv\varphi(T)/T$, the subscript specifying the temperature at which it is evaluated.  It is motivated by baryon number washout.  Inside a broken-phase bubble, the sphaleron energy scales schematically as
\begin{equation}
 E_{\rm sph}(T)\simeq\frac{4\pi\varphi(T)}{g_2}\,\mathcal{B}(T)
 ~,
 \label{eq:sphaleron-energy}
\end{equation}
where $\mathcal{B}$ is a shape factor of order one, and the washout rate is exponentially suppressed by $\exp[-E_{\rm sph}/T]$.  The relevant field value is the one at the nucleation temperature, $\xi_n=\varphi(T_n)/T_n$, not the degenerate critical value $\xi_c=\varphi(T_c)/T_c$.  In both cases, $\varphi$ is the value in the daughter phase measured from the origin, not the field discontinuity between the parent and daughter phases.  Both ratios are evaluated in a specified gauge.  Requiring the broken-phase washout rate to fall below the Hubble rate gives the rule of thumb $\xi_n\gtrsim1$.  The ratio $\xi_n$ is not an observable, and the rule stands in for a dedicated sphaleron calculation~\cite{KlinkhamerManton1984,Ahriche2014,Patel2011}.

\subsection{Conditions for a cosmological FOPT}
\label{subsec:ct-conditions}

Establishing a cosmological phase transition takes more than a pair of degenerate minima.  We apply, in sequence:
\begin{enumerate}
\item
The zero-temperature scalar spectrum must satisfy the theoretical, electroweak precision, Higgs, and direct-search bounds summarized in Sec.~\ref{sec:model-constraints}.

\item
The relevant phases must be resolved, and the global minimum at zero temperature must be the electroweak vacuum within the stated tolerance, with $\lambda_1$ fixed by the tree-level Higgs mass of Sec.~\ref{subsec:spectrum}.

\item
A first-order critical transition must belong to the resolved network of phases traced by CT and pairwise critical transitions.  It has a positive energy density difference $\rho_{\rm false}(T_c)-\rho_{\rm true}(T_c)$, which is the latent heat, where $\rho_i=V_i-T\,\partial V_i/\partial T$ for phase $i$ (`true' or `false').  This difference is nonzero even though the free energies are degenerate.

\item For the completed transition benchmarks, the action must converge over a temperature interval wide enough to integrate the bubble nucleation probability and test percolation and completion.  An evaluation that fails to converge remains unresolved.
\end{enumerate}

Dimensional reduction, which integrates out the nonzero Matsubara modes and organizes the remaining three-dimensional theory by scale~\cite{Niemi2018,Ekstedt2023}, is a complementary route to better perturbative control.  For a hypercharged complex triplet, though, its matching coefficients and higher-dimensional operators have to be derived for this model, which is beyond the scope of this paper.  We work instead in four dimensions at one loop with RG improvement.

\section{What the Critical Temperature Criterion Misses}
\label{sec:scan-results}

The broad survey in this section is used primarily to compare levels of phase transition analysis.  We fix $\lambda_3(Q_{\rm ref})=0$, which removes one coupling from the survey, and determine $\lambda_1(Q_{\rm ref})$ from the Higgs mass.  On the neutral plane, the tree-level potential depends on $\lambda_2$ and $\lambda_3$ only through $\lambda_2+\lambda_3$, although the boundedness and unitarity conditions depend on them separately.  At one loop, $\lambda_3$ enters the triplet thermal self-energy and the running of $M^2$, $16\pi^2\beta_{M^2}\supset12\lambda_3M^2$, and of $\lambda_2$, $\lambda_4$ and $\lambda_5$, and it is generated by the running even when it vanishes at $Q_{\rm ref}$.  Section~\ref{sec:completed-transition-benchmarks} comments on what a nonzero $\lambda_3$ does to the completed transitions.  The candidates, drawn from a Sobol sequence, which fills a box more evenly than independent random draws, cover
\begin{equation}
 0.05\leq\lambda_2\leq1.00
 ~,\qquad
 0.30\leq\lambda_4\leq4.25
 ~,\qquad
 -3.00\leq\lambda_5\leq2.99
 ~,
\end{equation}
and
\begin{equation}
 353\leq\frac{M_\Delta}{\mathrm{GeV}}\leq1000
 ~,
 \qquad
 1.0\times10^{-5}\leq\frac{v_\Delta}{\mathrm{GeV}}\leq1.98
 ~.
\end{equation}
The interval of $\lambda_5$ includes both signs and, thus, both orderings of the mostly triplet masses of Sec.~\ref{subsec:spectrum}.

A commonly used first pass labels a transition strong when the field norm on the low-temperature side of a critical degeneracy satisfies $v_{\rm low}(T_c)/T_c>1$.  The survey contains both one-step and multistep phase structures.  A one-step history connects the symmetric phase directly to the electroweak vacuum, while a multistep history first enters an intermediate broken phase and reaches the electroweak vacuum only through a later transition.  CT can also trace local branches that never become the occupied equilibrium phase.  The number of local minima or critical degeneracies is therefore not the number of cosmological transitions.

\begin{figure}[htbp]
\centering
\includegraphics[width=0.8\textwidth]{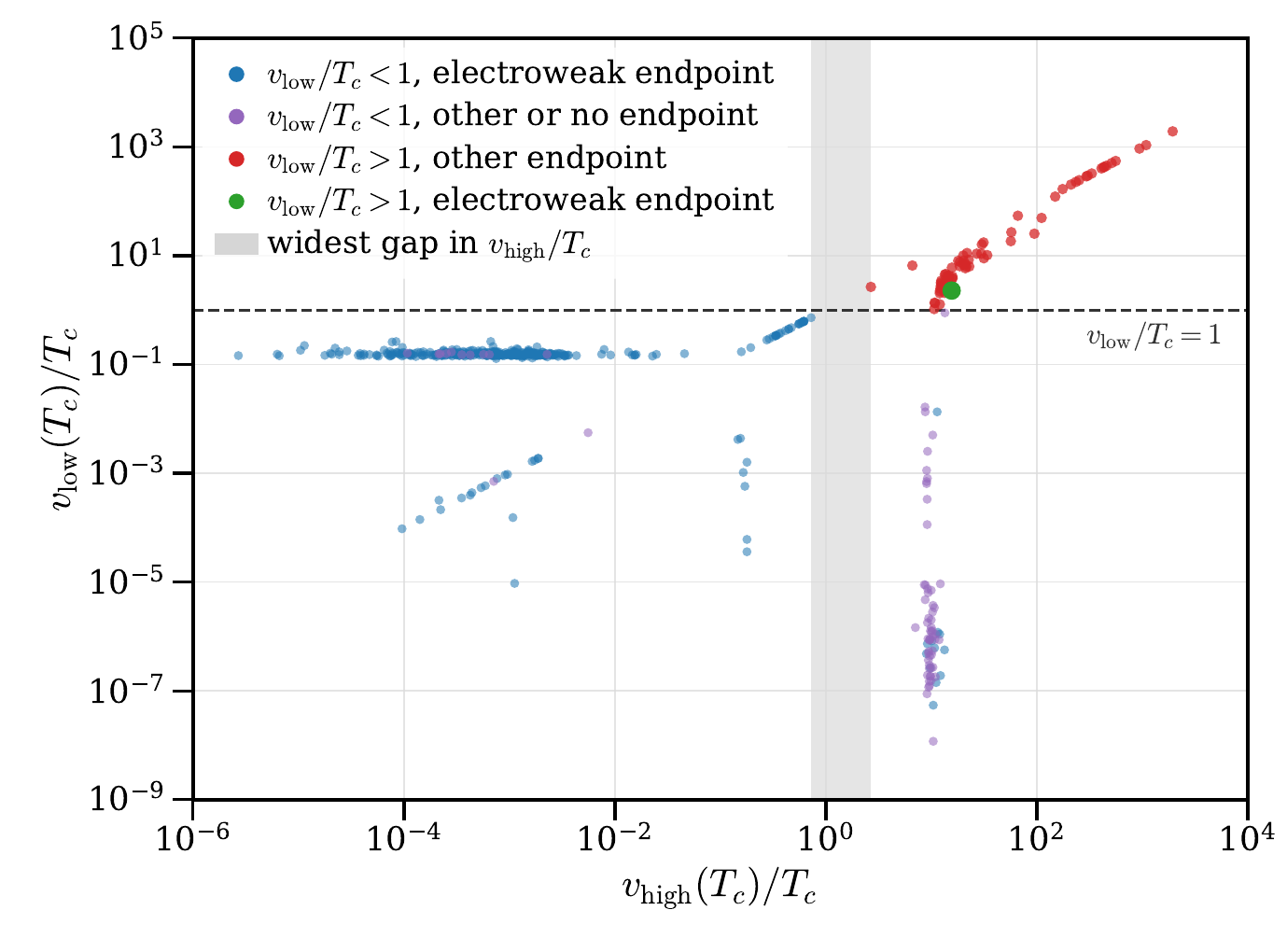}
\caption{Distribution of selected first-order critical transitions in the $(v_{\rm high}(T_c)/T_c, v_{\rm low}(T_c)/T_c)$ plane.  Blue and purple dots fail the cut $v_{\rm low}(T_c)/T_c>1$ and reach at zero temperature the required electroweak vacuum (blue) or another or no endpoint (purple).  Red dots pass the cut and reach another endpoint, and the single green dot passes it and reaches the electroweak vacuum.  The dashed line is $v_{\rm low}(T_c)/T_c=1$, and the gray band indicates the gap in $v_{\rm high}(T_c)/T_c$ over all of them.}
\label{fig:vc-over-tc-phase-graph}
\end{figure}

Figure~\ref{fig:vc-over-tc-phase-graph} shows what the cut $v_{\rm low}(T_c)/T_c>1$ actually selects.  The apparently strong candidates lie above the horizontal $v_{\rm low}(T_c)/T_c=1$ line, but every one of them has a high-temperature phase that is already broken rather than the electroweak-symmetric origin, and all but one of them evolve toward a zero-temperature endpoint other than the required electroweak vacuum.  The high-temperature field norm $v_{\rm high}(T_c)$ of every selected transition is several times $T_c$.  The widest gap in that ratio over all the FOPTs of the survey, the gray band, lies immediately below the smallest selected value.  Every transition the cut admits therefore has a strongly broken parent phase.  The converse does not hold.  A large value of $v_{\rm low}(T_c)/T_c$ alone does not establish a strong electroweak transition.

For baryon number preservation, the sphaleron suppression must be evaluated in the phase that holds the baryon asymmetry, after the electroweak breaking step.  For GW production, the strength and the tunneling action must belong to a transition that occurs in the cosmological history, starting from the phase the cooling Universe occupies.  Such a transition may also connect two broken phases.  A cut on $v_c/T_c$ imposes neither requirement.

The transition strength separates the selected candidates from the completed transitions, and at a critical degeneracy it needs no dynamical input.  The two phases have equal free energy, $\Delta V(T_c)=0$, but its temperature derivative does not vanish, so $\alpha_\theta$ of Eq.~\eqref{eq:gw-alpha-beta} reduces to
\begin{equation}
 \alpha_\theta(T_c)=-\frac{T_c}{4\rho_{\rm rad}}\left.\frac{d\Delta V}{dT}\right|_{T_c}
 =\frac{\Delta\rho(T_c)}{4\rho_{\rm rad}(T_c)}
 ~,
 \label{eq:alpha-at-tc}
\end{equation}
where $\Delta\rho=\rho_{\rm false}-\rho_{\rm true}$ is the latent heat of Sec.~\ref{subsec:ct-conditions}.  Setting aside $2$ entries with negative latent heat, which a cooling transition cannot have, the selected degeneracies have positive $\Delta\rho(T_c)$ and strengths running from about $2$ to $400$, that is, $5$ to $7$ orders of magnitude above the largest strength, $\alpha_\theta\simeq5\times10^{-5}$, among the $7$ completed transitions to be discussed in Sec.~\ref{sec:completed-transition-benchmarks} and not shown in Fig.~\ref{fig:vc-over-tc-phase-graph}.

Turning a strength into an amplitude needs the inverse duration, the wall speed and the nucleation temperature, none of which have been determined for the selected candidates, whose actions were not calculated.  The strength alone therefore bounds only the factor of the amplitude that depends on it, and no bound on the complete spectrum follows before those dynamical inputs are known.  At fixed inverse duration, wall speed, relativistic species count and transition temperature, the acoustic amplitude of Eq.~\eqref{eq:gw-components} carries $[\kappa_{\rm sw}\alpha_\theta/(1+\alpha_\theta)]^2$ for a long-lived source.  For a short-lived one, $H_*\tau_{\rm sw}\ll1$, the factor $\Upsilon\simeq H_*\tau_{\rm sw}\propto1/\overline U_f$ removes half a power and leaves the three-halves power of the same combination.  The peak frequency does not depend on the strength.  Holding $\kappa_{\rm sw}$ fixed understates the ratio, because $\kappa_{\rm sw}$ grows with $\alpha_\theta$.  The strength-dependent factor of the amplitude assigned to these candidates therefore exceeds that of the completed transitions by at least $6$ orders of magnitude, and by $8$ when the source is long-lived.

\section{Completed Phase Transitions and Their Phenomenology}
\label{sec:completed-transition-benchmarks}

Conditional on the plasma reaching the symmetric origin from higher temperatures, seven benchmark points in our sampled pool satisfy every condition imposed in this work: the zero-temperature conditions, the correlated $S,T,U$ condition at $95\%$ CL, the direct-search limits that apply at their triplet VEVs, and integrated nucleation, percolation, and completion, as described in the preceding sections.

The thermal histories and critical temperatures in this section are obtained using CT version~2.0.7~\cite{Wainwright2012,CosmoTransitions207}.  The bounce actions $S_3(T)$ that enter $T_n$ and $\beta/H_*$ are approximated by shooting along the straight path between the two minima.  The full two-field path deformation of CT on the same minima changes $\beta/H_*$ by at most $2\%$ and $T_n$ negligibly.  The nucleation, percolation, completion, and GW calculations use these actions only inside the temperature interval over which each is calculated.  Every result in this section fixes $m_h=125.13\,\mathrm{GeV}$, with $\lambda_1$ set at each parameter point by the tree-level $CP$-even mass matrix of Sec.~\ref{subsec:spectrum}.  Neither this condition nor the radial rematching of Eq.~\eqref{eq:radial-rematching} matches the one-loop pole mass.  Every retained point has a global zero-temperature minimum within $0.5$~GeV of the required electroweak vacuum and contains one direct FOPT from the symmetric phase at the origin to that vacuum, under the assumption on the earlier history stated in Sec.~\ref{subsec:completed-region-systematics}.

The field discontinuity $\Delta\varphi$ of the completed transition, the distance in field space between the two phases at $T_c$, always satisfies $\Delta\varphi/T_c\leq0.44$ in the central prescription.  The usual critical-strength proxy is therefore not required for a FOPT to complete.  As discussed in Sec.~\ref{sec:scan-results}, on its own it does not guarantee a successful transition either.

\subsection{Parameters of the completed points}
\label{subsec:completed-transition-cluster}

\begin{table}[htbp]
\centering
\setlength{\tabcolsep}{3.0pt}
\begin{tabular}{lrrrrrrr}
\toprule
Point & $\lambda_2$ & $\lambda_4$ & $\lambda_5$ &
\shortstack{$M_\Delta$\\(GeV)} & \shortstack{$v_\Delta$\\(MeV)} &
\shortstack{$m_{H^{\pm\pm}}$\\(GeV)} & $\chi^2$ \\
\midrule
$P_1$ & $0.717$ & $0.767$ & $2.816$ & $613.2$ & $379.8$ & $539.1$ & $7.797$ \\
$P_2$ & $0.372$ & $2.141$ & $1.187$ & $803.2$ & $0.7757$ & $780.5$ & $0.368$ \\
$P_3$ & $0.733$ & $1.039$ & $2.179$ & $638.1$ & $202.0$ & $584.1$ & $0.974$ \\
$P_4$ & $0.576$ & $1.153$ & $2.233$ & $710.7$ & $1406$ & $661.3$ & $0.127$ \\
$P_5$ & $0.470$ & $3.093$ & $-1.231$ & $876.0$ & $0.9492$ & $897.0$ & $0.134$ \\
$P_6$ & $0.976$ & $0.545$ & $2.322$ & $517.7$ & $100.1$ & $444.5$ & $6.604$ \\
$P_7$ & $0.597$ & $0.604$ & $2.648$ & $624.7$ & $23.55$ & $556.8$ & $4.828$ \\
\bottomrule
\end{tabular}
\medskip
\caption{The seven completed benchmark points, ordered by decreasing $\Delta\varphi/T_c$.  The five input parameters are evaluated at $Q_{\rm ref}=246.2$~GeV, $m_{H^{\pm\pm}}$ is the tree-level mass of Sec.~\ref{subsec:spectrum}, and $\chi^2$ is the correlated $S,T,U$ statistic against its $95\%$ limit.}
\label{tab:completed-benchmark-parameters}
\end{table}

Table~\ref{tab:completed-benchmark-parameters} gives the five input parameters of the seven benchmarks, the masses of their doubly charged Higgs bosons and their $\chi^2$ values for the oblique parameters, ordered by decreasing field discontinuity.  The points cover more than three orders of magnitude in the triplet VEV, from $v_\Delta \simeq 0.78$~MeV to $1.4$~GeV, and the doubly charged Higgs mass runs from about $440$ to $900$~GeV.  Six of them have $\lambda_5>0$ and therefore the mass ordering
\begin{equation}
 m_{H^{\pm\pm}}<m_{H^{\pm}}<m_{H^0}\simeq m_A
 ~.
 \label{eq:completed-benchmark-mass-ordering}
\end{equation}
At $P_5$, where $\lambda_5<0$, the ordering is reversed.  The doubly charged state is then the heaviest of the mostly triplet states, and $P_5$ is the only point with $m_{H^{\pm\pm}}>M_\Delta$.  With $\lambda_3=0$ on the sampled subspace, the doubly charged Higgs mass relation of Sec.~\ref{subsec:spectrum} is exact: $m_{H^{\pm\pm}}^2=M_\Delta^2-\lambda_5v_\Phi^2/2$.

One of the seven sits on the boundary of the electroweak fit.  The point $P_1$ has $\chi^2=7.797$ against the $95\%$ limit of $7.815$, and it has the strongest transition among the seven, with $\Delta\varphi/T_c=0.44$.  The strongest completed transition we find is thus the one closest to being removed by the electroweak data.  If a tighter oblique parameter fit excluded it, the largest value among the remaining six would be $\Delta\varphi/T_c=0.334$, at $P_2$.  The oblique parameters are evaluated with the tree-level masses of Sec.~\ref{subsec:spectrum}, as is consistent at one loop.

\begin{figure}[htbp]
\centering
\includegraphics[width=0.8\textwidth]{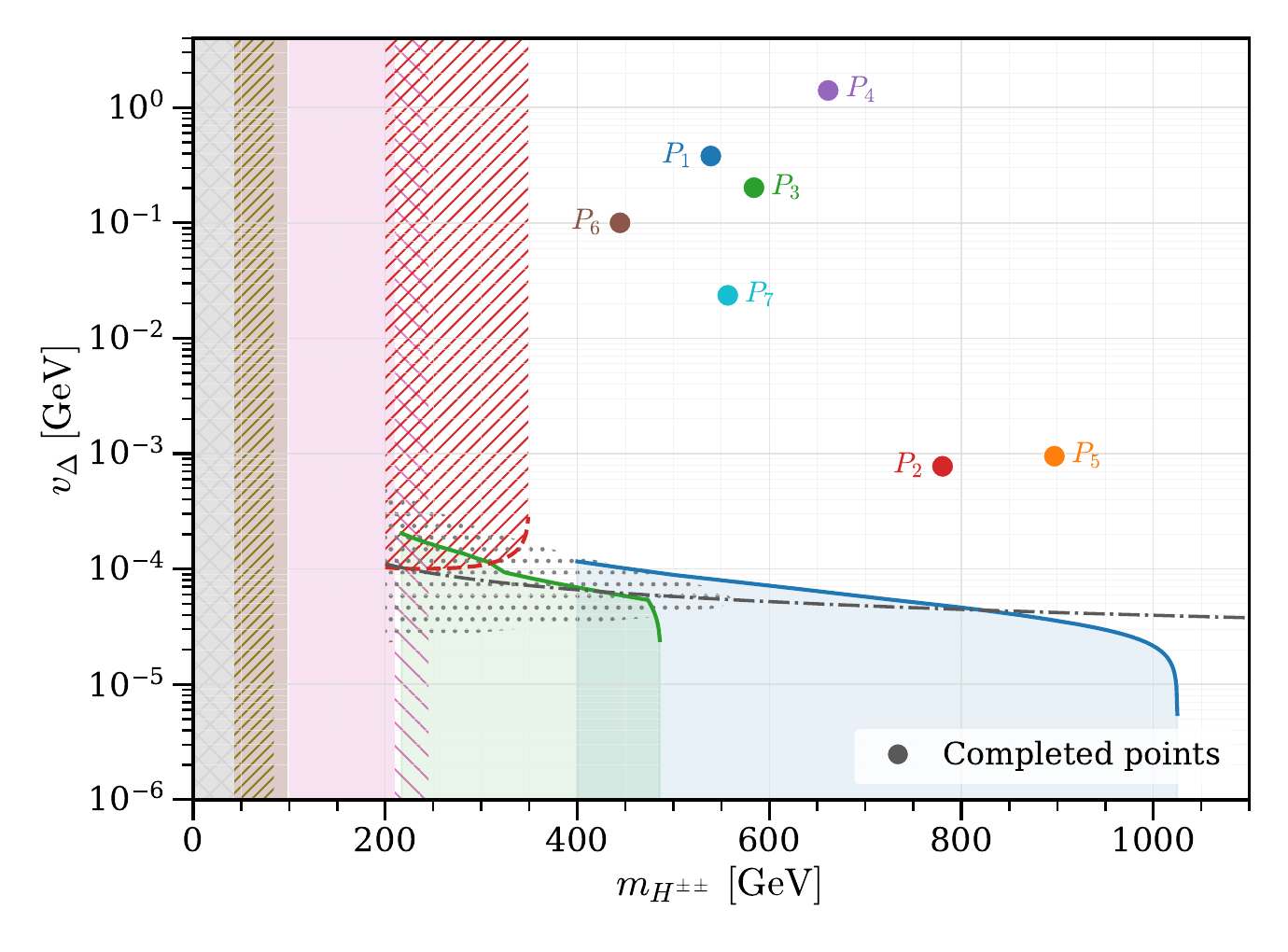}
\caption{Completed points in the $m_{H^{\pm\pm}}$--$v_\Delta$ plane, along with experimental search limits.  Solid blue and solid green curves, with the shaded areas below them, are the ATLAS prompt-dilepton exclusions at $139$ and $36.1$~fb$^{-1}$, and the dashed red curve with the hatched area above it is the ATLAS same-sign $WW$ exclusion.  The dash-dotted gray curve is the locus $\mathrm{BR}_{\ell\ell}=\mathrm{BR}_{WW}=1/2$, and the dotted area has $c\tau>10\,\mu\mathrm{m}$.  The vertical strips are, from the left, the LEP $Z$-width limit (gray, cross-hatched), the overlapping $8$~TeV $WW$ recast (olive, hatched) and OPAL dilepton (brown) limits, and the CDF prompt $e/\mu$ limit (pink) with its mode-dependent extension (pink, hatched).}
\label{fig:completed-cluster-hpp-vdelta}
\end{figure}

Figure~\ref{fig:completed-cluster-hpp-vdelta} places the seven points in the $m_{H^{\pm\pm}}$--$v_\Delta$ plane together with the direct-search limits.  Which limit applies at a given point is decided by $v_\Delta^{\rm eq}$, the triplet VEV where the same-sign lepton and diboson widths of Sec.~\ref{subsec:collider-constraints} are equal.  Over the mass range of the seven points, $v_\Delta^{\rm eq}$ is about $(4$--$6)\times10^{-5}$~GeV for the normal ordering with a massless lightest neutrino state, more than an order of magnitude below the smallest triplet VEV among them.  The dilepton widths are therefore negligible, and the prompt-dilepton limits do not constrain these points.  At the six points with $\lambda_5>0$ the doubly charged state is the lightest triplet state and decays to same-sign $WW$ pairs.  At $P_5$, it is the heaviest, about $10$~GeV above $H^\pm$, and the cascade $H^{++}\to H^+W^{+*}$ dominates its decays.  The same-sign $WW$ pair production limit applies at the six points with $\lambda_5>0$, and all of them lie above its $350$~GeV mass limit.  It does not apply at $P_5$.

The boundaries in the figure keep the decay and mass ordering assumptions of Sec.~\ref{subsec:collider-constraints}.  The prompt-dilepton limits of the ATLAS $139$ and $36.1$~fb$^{-1}$ searches~\cite{ATLAS2023Dilepton,ATLAS2018Dilepton} are translated into the $m_{H^{\pm\pm}}$--$v_\Delta$ plane with the normal-ordering best-fit mixing parameters of NuFIT~6.0~\cite{NuFIT2024} and a triplet Yukawa texture fixed by the neutrino mass matrix.  The same-sign $WW$ curve assumes a unit diboson branching fraction~\cite{ATLAS2021Charged}, which holds for the six completed points with $\lambda_5>0$ and not at $P_5$.

\subsection{Thermal properties of the completed points}
\label{subsec:completed-benchmark-physics}

\begin{table}[htbp]
\centering
\setlength{\tabcolsep}{4.0pt}
\begin{tabular}{lcccccc}
\toprule
Point & \shortstack{$T_c$\\(GeV)} & \shortstack{$T_n$\\(GeV)} & \shortstack{$(T_c-T_n)/T_c$\\($\%$)} &
$\Delta\varphi/T_c$ & \shortstack{$\alpha_\theta$ \\ $(\times 10^{-5})$} & \shortstack{$\beta/H_*$ \\ $(\times 10^5)$}  \\
\midrule
$P_1$ & ${358.3}_{-144.7}$ & $356.1$ & $0.61$ & ${0.435}_{-0.290}$ & $5.11$ & $0.4$ \\
$P_2$ & ${411.6}_{-184.4}$ & $410.4$ & $0.28$ & ${0.334}_{-0.211}$ & $3.40$ & $0.8$ \\
$P_3$ & ${320.8}_{-101.4}$ & $320.6$ & $0.07$ & ${0.238}_{-0.104}$ & $3.36$ & $3$ \\
$P_4$ & ${329.5}_{-103.9}$ & $329.3$ & $0.06$ & ${0.226}_{-0.097}$ & $3.26$ & $4$ \\
$P_5$ & ${420.7}_{-165.3}$ & $420.4$ & $0.07$ & ${0.204}_{-0.089}$ & $1.82$ & $3$ \\
$P_6$ & ${259.0}_{-61.0}$ & $258.9$ & $0.04$ & ${0.200}_{-0.058}$ & $3.37$ & $7$ \\
$P_7$ & ${249.1}_{-47.6}$ & $249.1$ & $0.02$ & ${0.180}_{-0.045}$ & $3.64$ & $10$ \\
\bottomrule
\end{tabular}
\medskip
\caption{Thermal properties of the seven completed points, with $\alpha_\theta$ and $\beta/H_*$ given by Eq.~\eqref{eq:gw-alpha-beta} at $T_*=T_n$.  The subscript value on $T_c$ and $\Delta\varphi/T_c$ is the largest downward shift over the common interval $0.50\leq\kappa\leq1.00$ of the renormalization scale factor $\kappa$ of Sec.~\ref{subsec:completed-region-systematics}.  That is the interval on which all seven points have a transition.  Above $\kappa=1$ the transition disappears at $P_1$ and $P_2$ and moves upward at the other five, as Table~\ref{tab:completed-region-scale-seven} shows.}
\label{tab:completed-representative-thermal}
\end{table}

Table~\ref{tab:completed-representative-thermal} collects the thermal properties of the seven completed points.  The strength $\alpha_\theta$ is evaluated at $T_*=T_n$.  Three features of the table matter.  First, both measures of strength are narrowly spread and uniformly small: $\Delta\varphi/T_c$ ranges from $0.18$ to $0.44$, and $\alpha_\theta$ stays within a factor of two of $3\times10^{-5}$ across points whose triplet VEVs differ by more than three orders of magnitude.  No point approaches the commonly used indicator $\Delta\varphi/T_c\gtrsim1$.  Second, the transitions are fast: $\beta/H_*$ lies between $4\times10^{4}$ and $1\times10^{6}$, so that each lasts between about $10^{-6}$ and $2.5\times10^{-5}$ of a Hubble time.  Finally, there is almost no supercooling: $T_n$ lies below $T_c$ by at most $0.6\%$, at $P_1$, and by as little as $0.02\%$, at $P_7$.

\begin{figure}[htbp]
\centering
\includegraphics[width=0.8\textwidth]{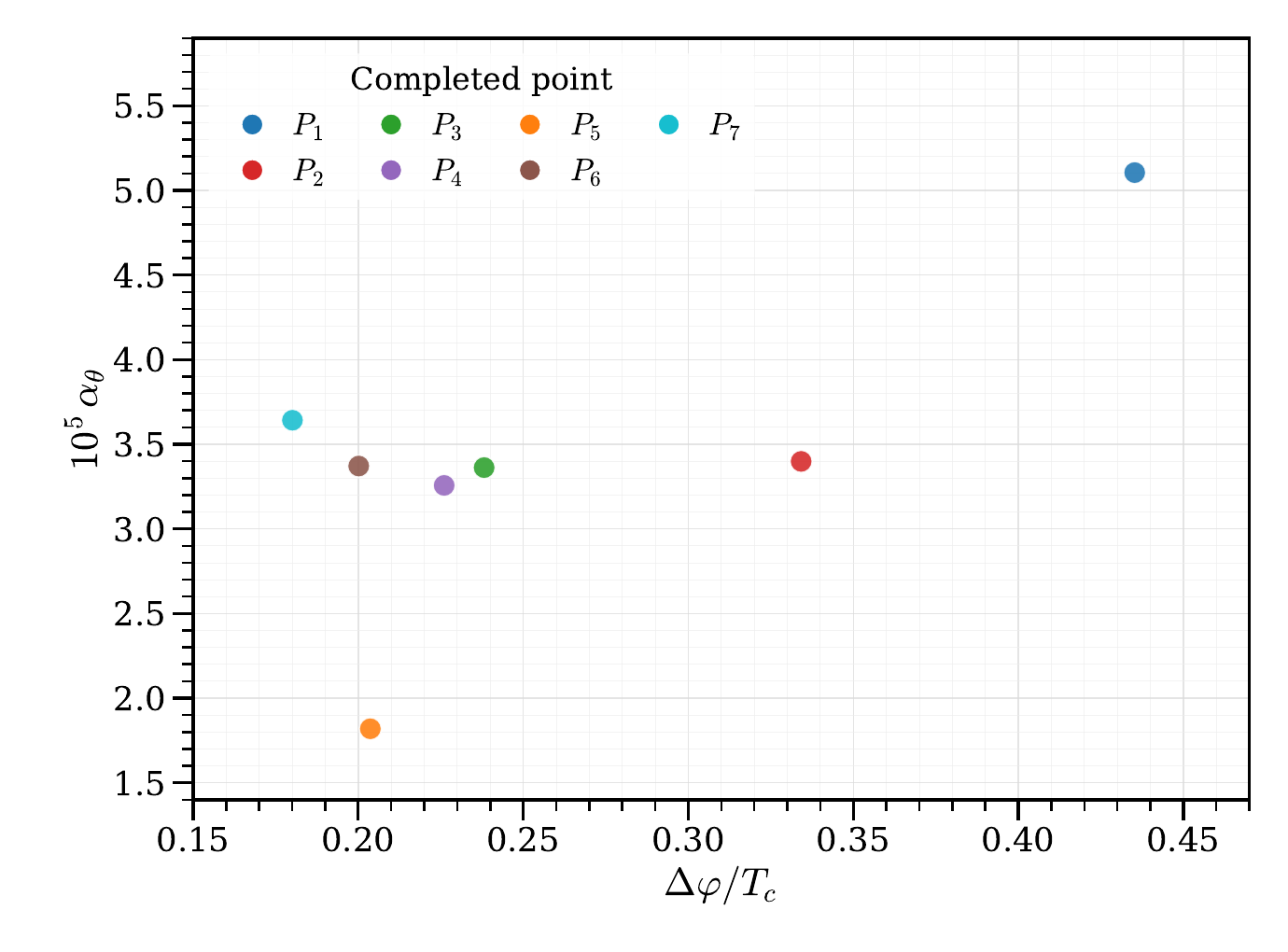}
\caption{Trace anomaly strength as $10^5\alpha_\theta$ against the field discontinuity $\Delta\varphi/T_c$ at the seven completed points.}
\label{fig:completed-region-strength-alpha}
\end{figure}

Two quantities are used for the strength of a transition.  The first is the field discontinuity $\Delta\varphi/T_c$, the distance in field space between the two phases at the critical temperature divided by that temperature.  The conventional $v_c/T_c$ is the field norm of the broken phase alone at $T_c$, and the two agree when the high-temperature phase sits at the exactly symmetric origin, as it does for the seven points retained here.  The second is the trace anomaly strength $\alpha_\theta$ of Eq.~\eqref{eq:gw-alpha-beta}, evaluated at $T_n$.  Figure~\ref{fig:completed-region-strength-alpha} shows that the two do not have any strong correlation over the completed benchmarks found here.  Both span a factor of two to three, and $P_1$ is the largest in both, but below it their orderings differ.  There is no reason for them to agree.  The field discontinuity is a property of the potential at one temperature, while $\alpha_\theta$ is built from the free-energy difference between the phases and its temperature derivative.  A transition can move the field a long way while releasing little energy, or the reverse.  Of the two, only $\alpha_\theta$ enters the GW amplitude.

A low and narrow barrier is consistent with the second and third features.  In the thin-wall limit $S_3\simeq16\pi\sigma^3/(3\Delta V^2)$, with a wall tension $\sigma$ set by the barrier and $\Delta V$ growing roughly linearly in $T_c-T$ just below $T_c$.  A small $\sigma$ lets the nucleation condition be met soon after the two phases stop being degenerate.  The result is little supercooling and a steep $d(S_3/T)/dT$ at $T_n$, i.e., a large $\beta/H_*$.  Neither follows from the barrier height alone, since the action also depends on $\Delta V$.  The rankings of the barrier height at $T_c$, in units of $T_c^4$, and of the field discontinuity are identical, whereas those of barrier height and $\beta/H_*$ are nearly reversed.  The small $\alpha_\theta$ does not follow from the barrier, as Fig.~\ref{fig:completed-region-strength-alpha} already shows.

The electroweak vacuum reached by these transitions stays doublet dominated: the largest zero-temperature triplet VEV among the seven is only $1.4$~GeV, at $P_4$.  The triplet states are not light compared with the plasma: $M_\Delta/T_c$ runs from $1.7$ to $2.5$, where Fig.~\ref{fig:tadpole} shows the massless screening coefficient to be several times the exact one.  The ring term of the central prescription therefore overstates the triplet contribution to the screening of the zero modes.

The strength is the least stable of these quantities.  Its numerator, $\Delta V-(T/4)\,d\Delta V/dT$ with both phases at $T_n$, is a small residue of much larger one-loop terms.  The vacuum energy term $\Omega$ is field-independent, so it enters the difference between the phases only because each phase sits at its own renormalization scale.  Since $\Omega$ runs at one loop, removing it is a one-loop change, and it raises the numerator by a factor of $4$ to almost $100$.  Removing the radial-rematching term of Eq.~\eqref{eq:radial-rematching}, which is formally of higher order, changes the numerator by up to a factor of $25$ and reverses its sign at all seven points.  The magnitude of $\alpha_\theta$ is therefore known only to within one to two orders of magnitude, and its sign is not stable under the radial rematching diagnostic.  Even with the largest of these values, the sound wave peaks of Sec.~\ref{subsec:completed-benchmark-gw} stay more than $5$ orders of magnitude below the deepest sensitivity curve.

The nucleation temperature is obtained by integrating the nucleation rate over the stored actions, while the inverse duration is the logarithmic slope $\beta/H_*=T_n\,d(S_3/T)/dT|_{T_n}$, fitted to $\log(S_3/T)$ over a window of converged action samples around $T_n$.  It is the only entry of Table~\ref{tab:completed-representative-thermal} determined by a fit, and the least well resolved.  A second evaluation of the slope, by a finite-difference stencil on the same action samples, differs from the fitted value by $2\%$ to $7\%$.  The acoustic amplitude scales as $(\beta/H_*)^{-2}$ for a short-lived source, $H_*\tau_{\rm sw}\ll1$, through its explicit factor and through the lifetime factor $\Upsilon$.  This holds approximately at every point for $v_w=0.6$ and at $P_3$ to $P_7$ for $v_w=0.95$, where the power lies between $-1.6$ and $-2$, so the amplitude carries $1.6$ to $2$ times the fractional uncertainty of $\beta/H_*$.  At $P_1$ and $P_2$ for $v_w=0.95$, where $H_*\tau_{\rm sw}\simeq1$ and $\Upsilon\simeq0.4$ to $0.5$, the power lies between $-1$ and $-2$.  The spectra below use the fitted central value and do not show this variation.

\subsection{\texorpdfstring{Sphaleron washout at nucleation}{Sphaleron washout at nucleation}}
\label{subsec:sphaleron-at-nucleation}

The sphaleron criterion of Sec.~\ref{subsec:sphaleron-criterion} is evaluated on the daughter phase field value $\xi_n=\varphi(T_n)/T_n$.  None of the seven completed points meet it at the common shape factor adopted here.  Their $\xi_n$ ranges from $0.19$ to $0.46$, the largest at $P_1$.  With $4\pi/g_2$ and the shape factor $\mathcal{B}=1.9$ of Eq.~\eqref{eq:sphaleron-energy}, they give $E_{\rm sph}(T_n)/T_n$ between $7$ and $17$, against about $35$, the value of $\log[T/H(T)]$ obtained by equating the broken-phase washout rate, written as $T\exp(-E_{\rm sph}/T)$ with a unit prefactor, to the Hubble rate.  At that shape factor, the threshold corresponds to $\xi_n\simeq1$.  Even the largest of the seven falls short by a factor of about two and the smallest by about five.  Baryon number preservation is thus not supported at any of the completed points within this estimate.  This shortfall is larger than the spread of $\mathcal{B}$ over the $\mathcal{O}(1)$ range in a sphaleron profile calculation, but a point-by-point sphaleron calculation would be needed for a definitive statement.  This concerns washout in the broken phase and does not enter the GW signal.  At these points, $\xi_n$ exceeds $\Delta\varphi/T_c$ by $5\%$ to $6\%$.  The parent phase is the symmetric origin, and the broken minimum deepens quickly just below $T_c$, so even a supercooling of a fraction of a percent raises the broken field by this much.

\subsection{\texorpdfstring{Renormalization scale dependence}{Renormalization scale dependence}}
\label{subsec:completed-region-systematics}

The renormalization scale was varied at all the points of Table~\ref{tab:completed-benchmark-parameters}, over the $21$ values of $\kappa=Q_{\rm ren}/Q(\varphi,\delta,T)$ from $0.50$ to $1.50$ in steps of $0.05$, with $Q$ the running scale of Eq.~\eqref{eq:central-running-scale}.  The construction has already been described in Secs.~\ref{sec:one-loop-rg} and~\ref{sec:thermal-potential}.

Two tests are applied independently at each scale choice.  The first is whether the exactly symmetric origin $\varphi=\delta=0$ is a local minimum over the high-temperature interval containing $T_c$, and is evaluated at the origin itself rather than on the nearest traced branch.  The second asks whether the direct FOPT from that origin to the broken vacuum is present.

\begin{figure}[tbp]
\centering
\begin{minipage}[t]{0.49\textwidth}
\centering
\includegraphics[width=\linewidth]{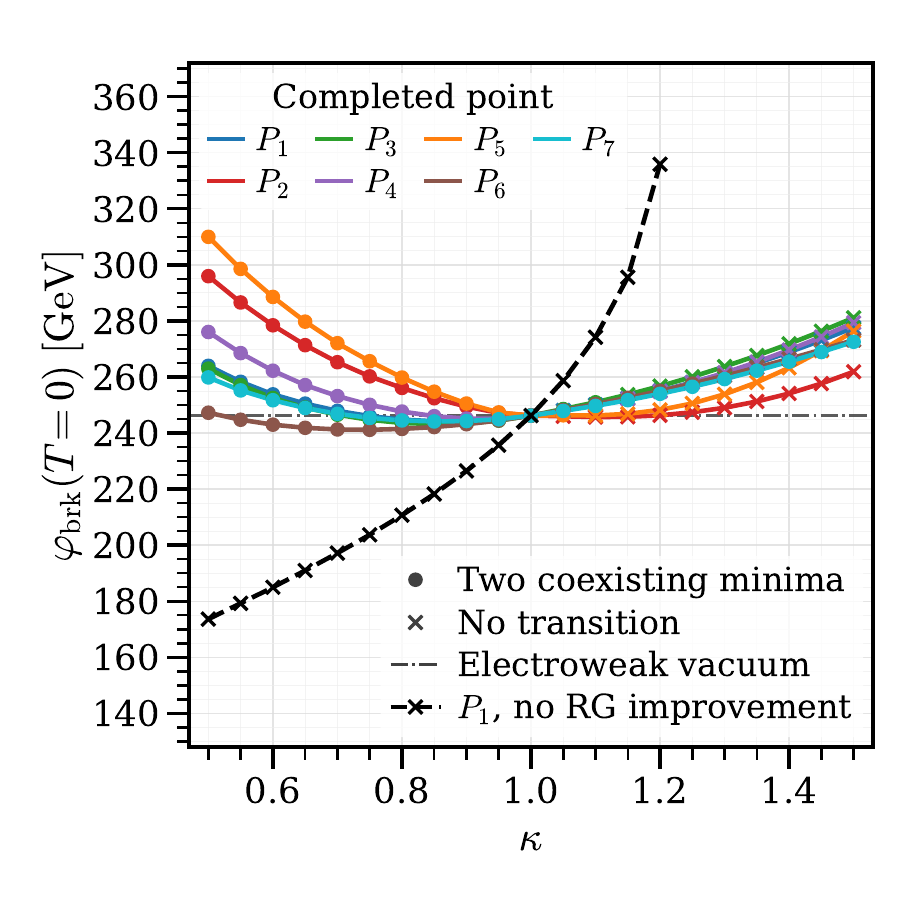}
{\small (a)}
\end{minipage}\hfill
\begin{minipage}[t]{0.49\textwidth}
\centering
\includegraphics[width=\linewidth]{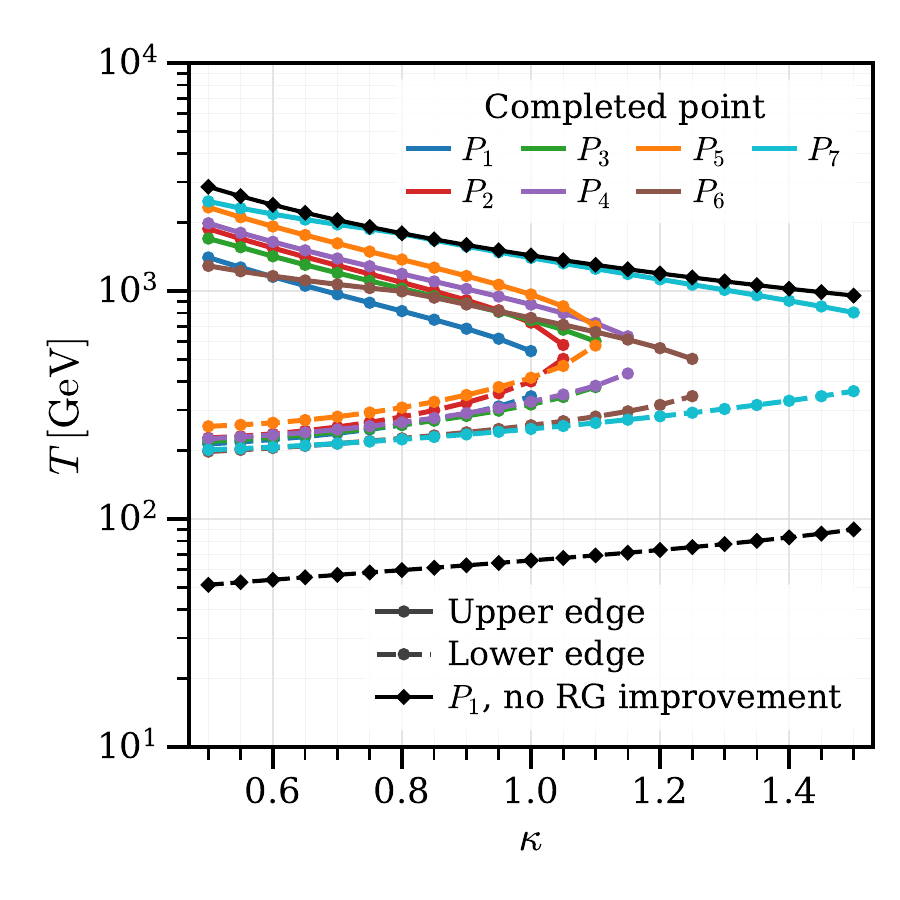}
{\small (b)}
\end{minipage}
\caption{(a) Broken-phase field magnitude at zero temperature, $\varphi_{\rm brk}(T=0)=\sqrt{\varphi^2+2\delta^2}$ in GeV, against the renormalization scale factor $\kappa$, at the seven completed points and all $21$ scale choices.  A dot marks a scale choice whose critical temperature comes from two coexisting local minima, and a cross marks one with no transition.  For contrast, the black dashed curve repeats $P_1$ without RG improvement and stops below $\kappa=1.25$.  The dash-dotted horizontal line marks the electroweak vacuum, $v=246.2$~GeV.  (b) Both edges of the high-temperature interval over which the exactly symmetric origin $\varphi=\delta=0$ is a local minimum against the renormalization scale factor $\kappa=Q_{\rm ren}/Q(\varphi,\delta,T)$, at the seven completed points, one color per point.  Solid curves are the upper edge and dashed curves are the lower edge.  Black diamonds repeat $P_1$ without RG improvement.}
\label{fig:completed-region-scale-fields}
\label{fig:completed-region-origin-interval}
\end{figure}

The scale variation moves the zero-temperature vacuum, as shown in Fig.~\ref{fig:completed-region-scale-fields}(a) for the completed transitions.  Away from the central scale, the finite-order potential describes a transition into a vacuum that is not the measured one.  At every benchmark, the departure is smallest at $\kappa=1$, where it is below $0.1$~GeV, and it reaches tens of GeV at the upper end of the grid at every point and at the lower end at all but $P_6$, which is the behavior of a truncated series whose counterterms were fixed at one scale.

Under the scale variation, the transition ends where the symmetric phase ends, or one grid step before.  Figure~\ref{fig:completed-region-origin-interval}(b) shows that as $\kappa$ grows, the lower edge of the high-temperature interval over which the exactly symmetric origin is a local minimum rises and the upper edge falls, so the high-temperature interval that can hold the transition ends at some value of $\kappa$.  At every point except $P_7$, its two edges meet inside the grid and the curves stop there.  Only $P_7$ keeps a wide interval over the entire grid, which is why it alone carries a transition out to $\kappa=1.50$.  A scale sensitivity has to be quoted together with the upper limit of its interval.

The upper edge also bounds the thermal history from above.  At $\kappa=1$ it lies between about $0.5$ and $1.4$~TeV, and there the curvature of the potential at the origin along $\varphi$ turns negative.  Along $\varphi$, the positive thermal contribution and the negative ring contribution nearly cancel at the upper edge.  Running $\lambda_4$ and $\lambda_5$ to $Q\simeq\pi T$, which also changes the triplet thermal self-energy built from them, strengthens the negative triplet scalar ring contribution until the remaining contributions no longer leave a positive curvature.  The field-dependent scale of Eq.~\eqref{eq:central-running-scale} adds the smaller negative term $(\partial V/\partial Q)\,\partial^2Q/\partial\varphi^2$, nonzero at the origin although $\partial Q/\partial\varphi$ vanishes there, and lowers the upper edge by about $100$~GeV.  With the couplings frozen, the upper edge is absent.  With the running couplings, above the edge, the minimum of the potential is a broken phase with $\varphi\simeq(0.7$--$0.8)\,T$ and a small $\delta$, and this phase is already deeper than the origin from $20$ to $120$~GeV below the edge.  The RG-improved potential does not restore the electroweak symmetry at temperatures above the edge.  The direct transitions of this section therefore assume that the plasma reaches the symmetric origin before cooling through $T_c$.

\begin{figure}[htbp]
\centering
\begin{minipage}[t]{0.49\textwidth}
\centering
\includegraphics[width=\linewidth]{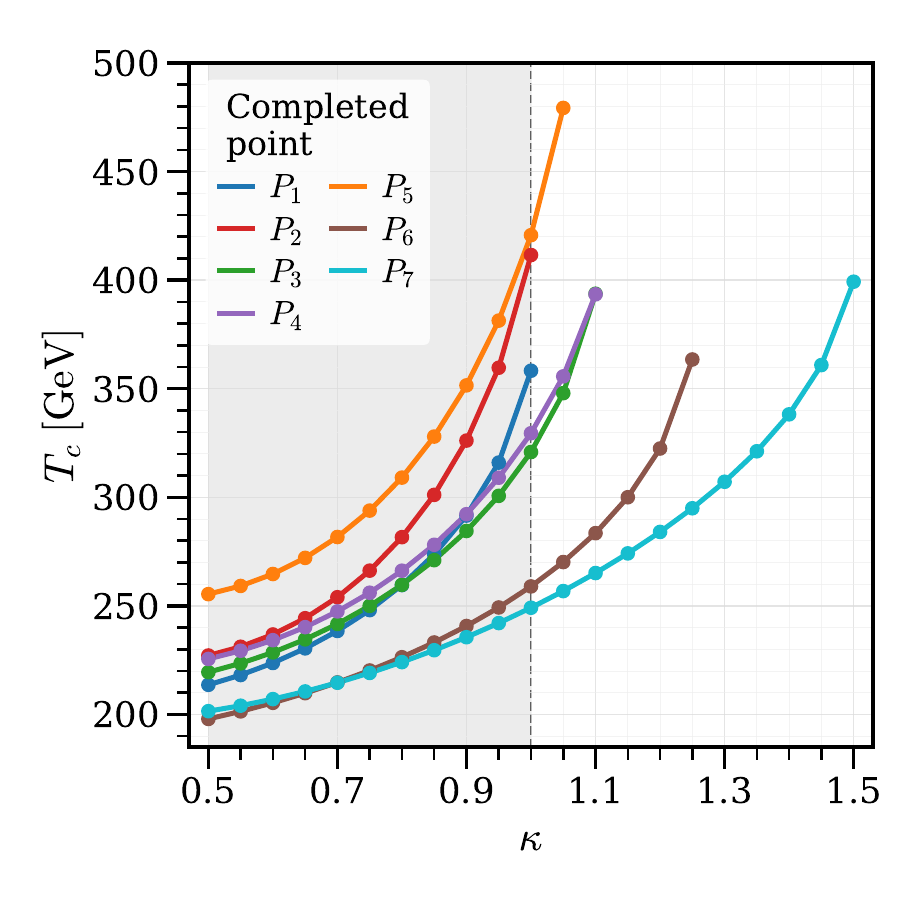}
{\small (a)}
\end{minipage}\hfill
\begin{minipage}[t]{0.49\textwidth}
\centering
\includegraphics[width=\linewidth]{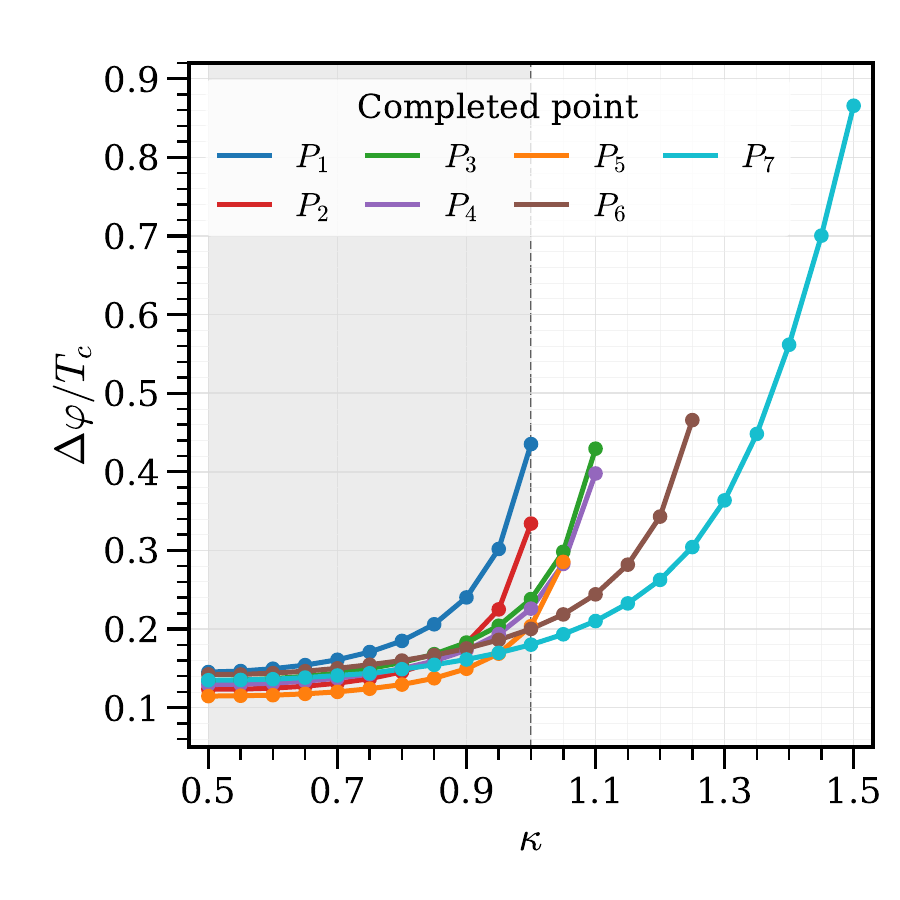}
{\small (b)}
\end{minipage}
\caption{Scale dependence of (a) the critical temperature $T_c$ and (b) the dimensionless ratio $\Delta\varphi/T_c$, against the renormalization scale factor $\kappa$, for the direct symmetric-to-broken FOPT at the seven completed points.  Each curve ends where its transition ends.  The dashed vertical line is $\kappa=1$ and the shaded band is $0.50\leq\kappa\leq1.00$, the range of $\kappa$ over which the critical temperature of each transition comes from two coexisting local minima.}
\label{fig:completed-region-scale-transition-observables}
\end{figure}

Figure~\ref{fig:completed-region-scale-transition-observables} shows that for all seven benchmark points, both $T_c$ and $\Delta\varphi/T_c$ grow with $\kappa$, slowly near $\kappa=0.5$ and steeply toward the end of each curve.  Except at $P_7$, which keeps its transition to $\kappa=1.50$, each curve ends at, or one grid step before, the scale where the high-temperature origin interval of Fig.~\ref{fig:completed-region-origin-interval}(b) ends.  The strongest transitions thus sit at the edge of the scale range where a transition exists at all.  There the zero-temperature vacuum has already moved away from the measured one, by $26$~GeV for the largest value, $\Delta\varphi/T_c=0.87$ at $P_7$ and $\kappa=1.50$.  On the shaded band where all seven points have a transition, $\Delta\varphi/T_c$ stays between $0.11$ and $0.44$.  The choice of scale changes the strength by a factor of a few, but none on the grid brings $\Delta\varphi/T_c$ to $1$.

The black curves of Fig.~\ref{fig:completed-region-scale-fields} repeat $P_1$ with the RG improvement switched off.  At none of the $21$ scale choices does that potential give a FOPT into a stable vacuum.  Its only first-order degeneracy, at $\kappa=1.25$ with $T_c\simeq75$~GeV and $\Delta\varphi/T_c\simeq0.09$, occurs where the zero-temperature potential has no minimum near $v$.  At the other scale choices, symmetry restoration is continuous: the broken minimum walks into the origin at the temperature where the origin stops being a maximum.  This is why Fig.~\ref{fig:completed-region-scale-transition-observables} carries no unimproved curve, since neither a critical temperature nor a field discontinuity exists to draw.  This absence was checked by continuing the broken minimum independently in temperature.  Away from $\kappa=1.25$, it ends where it merges into the origin, without any degeneracy.  The upper edge drawn for that point in Fig.~\ref{fig:completed-region-origin-interval}(b) is the top of the scale domain.  The curve of Fig.~\ref{fig:completed-region-scale-fields}(a) stops because that potential has no zero-temperature minimum at and beyond $\kappa=1.25$.  We show $P_1$ only.  Without the improvement, $P_2$, $P_4$ and $P_5$ restore the symmetry continuously at every $\kappa$, and $P_3$, $P_6$ and $P_7$ have only weak degeneracies, $\Delta\varphi/T_c\leq0.12$.

The two treatments can still be compared where both quantities exist, and they differ in opposite ways.  With the improvement, the broken field at zero temperature stays within about $30$~GeV of the vacuum over the $\kappa$ variation, while without it the field ranges from $174$ to $336$~GeV and no minimum exists at or above $\kappa=1.25$.  The temperature at which the origin stops being a maximum goes the other way: over $0.50\leq\kappa\leq1.00$ it varies about twice as much in relative terms with the improvement as without, $62\%$ against $28\%$, and at $\kappa=1$ it is $345$~GeV with the improvement against $66$~GeV without.  The two prescriptions do not differ only by terms of higher order.  Without the improvement, the parameters stay at $Q_{\rm ref}$ while the explicit logarithms are taken at $Q_{\rm ren}$, so nothing cancels the explicit scale dependence of the one-loop terms.  The two potentials then differ already at one loop, by $\log(Q_{\rm ren}/Q_{\rm ref})$ times the running of the tree potential, $\sum_X\beta_X\partial_XV_0-\gamma_\Phi\varphi\partial_\varphi V_0-\gamma_\Delta\delta\partial_\delta V_0$.  At $P_1$ and $\kappa=1$, near the critical temperature $T_c=358$~GeV of the RG-improved potential, that logarithm is about $1.5$.  The comparison therefore shows how much of the phase history is affected by the resummed logarithms.

\begin{table}[htbp]
\centering
\setlength{\tabcolsep}{3.5pt}
\begin{tabular}{lccc}
\toprule
Point &
\shortstack{Inside\\$\kappa$ window} &
\shortstack{$T_c$ (GeV) over\\the window} &
\shortstack{$\Delta\varphi/T_c$ over\\the window} \\
\midrule
$P_1$ & $0.50$--$1.00$ & $213.6$--$358.3$ & $0.145$--$0.435$ \\
$P_2$ & $0.50$--$1.00$ & $227.1$--$411.6$ & $0.123$--$0.334$ \\
$P_3$ & $0.50$--$1.10$ & $219.4$--$393.7$ & $0.134$--$0.429$ \\
$P_4$ & $0.50$--$1.10$ & $225.6$--$393.5$ & $0.129$--$0.398$ \\
$P_5$ & $0.50$--$1.05$ & $255.4$--$479.3$ & $0.115$--$0.285$ \\
$P_6$ & $0.50$--$1.25$ & $197.9$--$363.5$ & $0.142$--$0.466$ \\
$P_7$ & $0.50$--$1.50$ & $201.5$--$399.3$ & $0.135$--$0.866$ \\
\bottomrule
\end{tabular}
\medskip
\caption{Scale variation at the seven completed points, over the $21$-point grid $0.50\leq\kappa\leq1.50$ with RG improvement throughout.  The second column is the range of $\kappa$ over which the critical temperature lies inside the interval where the exactly symmetric origin is a local minimum, and the last two give $T_c$ and $\Delta\varphi/T_c$ over that range.}
\label{tab:completed-region-scale-seven}
\end{table}

Table~\ref{tab:completed-region-scale-seven} collects the outcome at each point.  Every window contains the band $0.50\leq\kappa\leq1.00$, shaded in Fig.~\ref{fig:completed-region-scale-transition-observables}, on which each of the seven points has a transition between two coexisting local minima, with one of them being the origin at the critical temperature.  From $\kappa=0.50$ to $\kappa=1.00$, $T_c$ rises by $20\%$ to $80\%$ and $\Delta\varphi/T_c$ by a factor of $1.3$ to $3$.  The benchmark $P_7$ keeps its transition over the entire grid.  At the other six points, the transition is lost above the upper end of the window in one of two ways.  At $P_1$, $P_3$ and $P_6$, the high-temperature interval of Fig.~\ref{fig:completed-region-origin-interval}(b) ends at the next grid step, so there is no symmetric phase to leave.  At $P_2$, $P_4$ and $P_5$, the interval survives one step further, but at that step no two phases are degenerate anywhere, and at the next step it ends as well.  Only the latter, at these three scale choices, is an absence in the sense of the conditions of Sec.~\ref{subsec:ct-conditions}.  At every other scale choice above a window, the symmetric phase itself does not survive the scale change, a property of the potential and not a limit of the traced temperature range.

Finally, we briefly comment on the effect of the top Yukawa coupling.  Taking it from the pole mass directly, $y_t=\sqrt{2}\,M_t/v$, instead of the $\overline{\mathrm{MS}}$ coupling of Sec.~\ref{subsec:field-content} moves $T_c$ down by $7\%$ to $19\%$, and a FOPT remains at all seven points.

\subsection{GW implications}
\label{subsec:completed-benchmark-gw}

Under the assumption that the walls are driven by the plasma and do not run away, as stated in Sec.~\ref{subsec:gw-spectra}, no scalar-field collision contribution is included.  The choice $\epsilon_{\rm turb}=0.10$ is illustrative, so the acoustic and turbulent spectra are given separately below.

\begin{figure}[htbp]
\centering
\begin{minipage}[t]{0.49\textwidth}
\centering
\includegraphics[width=\linewidth]{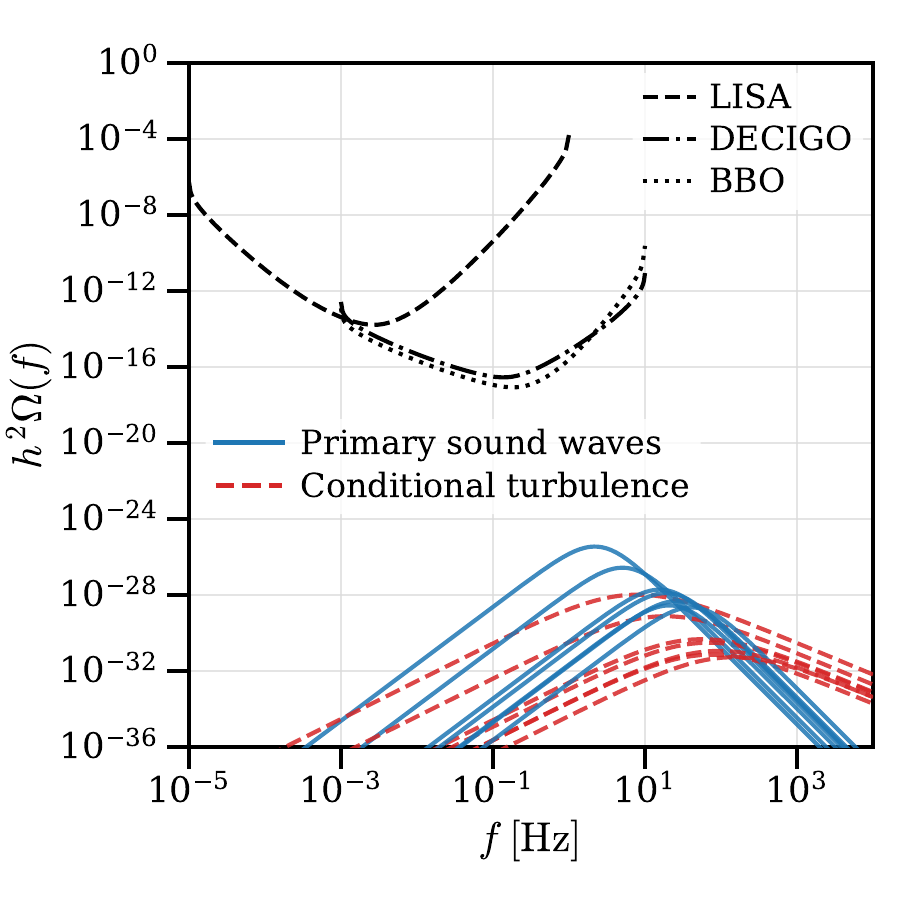}
{\small (a)}
\end{minipage}\hfill
\begin{minipage}[t]{0.49\textwidth}
\centering
\includegraphics[width=\linewidth]{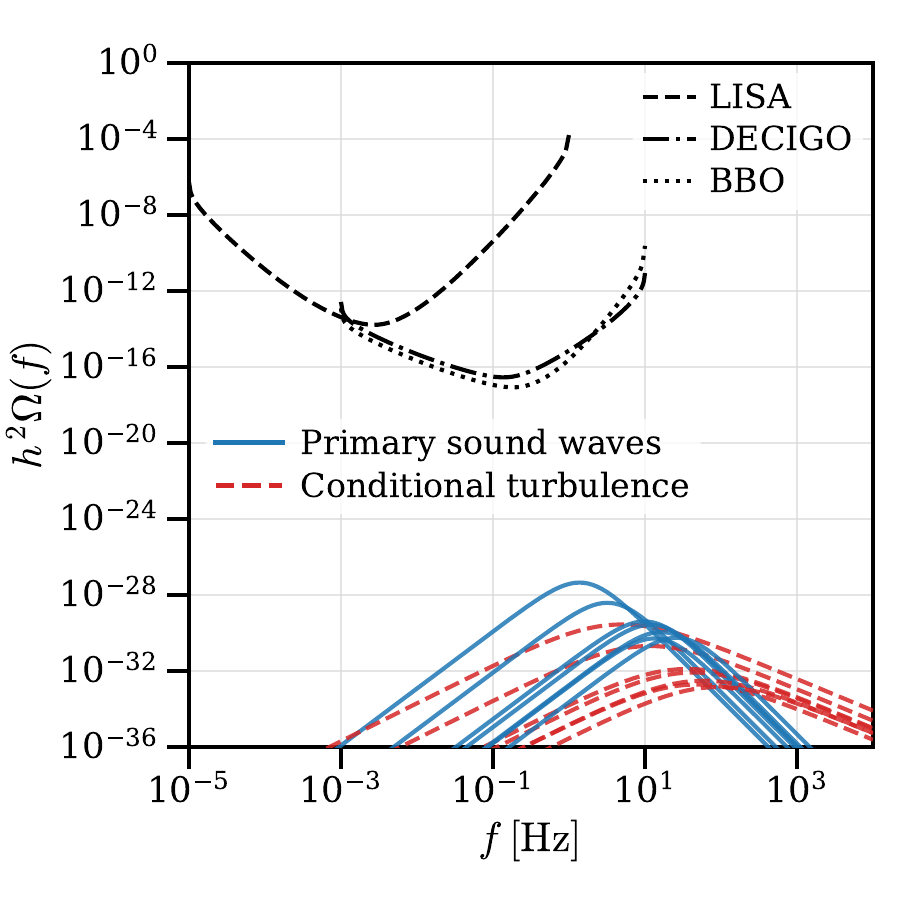}
{\small (b)}
\end{minipage}
\caption{GW energy density $h^2\Omega(f)$ against present-day frequency $f$ at wall speeds (a) $v_w=0.6$ and (b) $v_w=0.95$, for the seven completed transitions.  Blue solid curves are the sound wave spectra, and red dashed curves are the turbulent spectra at $\epsilon_{\rm turb}=0.10$.  Black curves are one-year power-law-integrated sensitivities at a signal-to-noise ratio of one~\cite{Schmitz2021}.}
\label{fig:completed-gw-spectra}
\end{figure}

The numbers in the last two columns of Table~\ref{tab:completed-representative-thermal} set the size and shape of the spectrum.  Setting $\kappa_{\rm sw}\leq1$, $\Upsilon\leq1$ and $v_{\rm sh}\leq1$ in Eq.~\eqref{eq:gw-components}, and inserting the largest trace anomaly strength and the smallest inverse duration of the sample, $\alpha_\theta=5.1\times10^{-5}$ and $\beta/H_*=4\times10^{4}$, both at $P_1$, bound the acoustic peak by
\begin{equation}
 h^2\Omega_{\rm sw,peak}
 <2.59\times10^{-6}\,
 \frac{1}{\beta/H_*}
 \left(\frac{\alpha_\theta}{1+\alpha_\theta}\right)^{2}
 \left(\frac{100}{g_*}\right)^{1/3}
 \simeq2\times10^{-19}
 ~.
 \label{eq:gw-sw-bound}
\end{equation}
Every factor dropped in reaching this inequality is smaller than one, and the fluid efficiency $\kappa_{\rm sw}$ at $\alpha_\theta\sim10^{-5}$ is far below unity.  The true peaks therefore lie well under the bound, which in the central prescription is already orders of magnitude below the power-law-integrated sensitivities of LISA, DECIGO and BBO in Fig.~\ref{fig:completed-gw-spectra}.  The peak frequency, from Eq.~\eqref{eq:gw-frequencies}, is set by the product of $\beta/H_*$ and $T_*$, and with $\beta/H_*\geq4\times10^{4}$ and $T_*\gtrsim250$~GeV it lies far above the millihertz band usually associated with a slow electroweak transition.  A low and narrow barrier is consistent with the small supercooling and the large $\beta/H_*$, while the small $\alpha_\theta$ is a separate feature of these points, as Sec.~\ref{subsec:completed-benchmark-physics} shows.  Together, they push the amplitude down and the frequency up at the same time.

Both inputs to Eq.~\eqref{eq:gw-sw-bound} come from the same point, $P_1$, which is the one retained by the electroweak fit by the smallest amount.  Without it the bound tightens to $4\times10^{-20}$, about four times lower, the largest strength then coming from $P_7$ and the smallest inverse duration from $P_2$.  The bound falls because the strength enters squared and the inverse duration inversely, and the remaining points have both a smaller strength and a shorter duration.

The limit of Eq.~\eqref{eq:gw-sw-bound} sets the fluid efficiency and the acoustic lifetime suppression to unity.  The fluid efficiency is far below unity at every point and enters squared, and $\Upsilon$ is at most $0.46$.  Hence, the computed peaks of Table~\ref{tab:completed-benchmark-gw-peaks} lie further below the sensitivity curves than the limit does.  The inverse duration tracks the supercooling.  With $\epsilon=(T_c-T_n)/T_c$ and $S_3/T\propto(T_c-T)^{-p}$ near $T_n$, Eq.~\eqref{eq:gw-alpha-beta} gives $\beta/H_*=p\,(1-\epsilon)(S_3/T)_n/\epsilon$.  The action at nucleation, $(S_3/T)_n=132$ to $135$, is fixed by $N(T_n)=1$.  With the prefactor of Eq.~\eqref{eq:thermal-nucleation-rate}, $\Gamma/H^4=1$ would instead require $S_3/T\simeq145$ to $147$ at these temperatures.  Because the action rises steeply above $T_n$, the integrated condition amounts approximately to $\Gamma/H^4\simeq\beta/H_*$ at $T_n$, lowering the required action by $\log(\beta/H_*)\simeq11$ to $14$.  The local exponent is $p=1.7$ to $1.8$ at $T_n$, rising toward the thin-wall value $2$.  The product $(\beta/H_*)\,\epsilon$ is therefore nearly the same at every point.  At fixed $(S_3/T)_n$ and $p$ the duration in Hubble units is $H_*/\beta=\epsilon/[p(1-\epsilon)(S_3/T)_n]\propto\epsilon$, so a point with more supercooling has a longer transition.  The transitions are short here because the supercooling is small at every one of the seven points, at most $0.6\%$.

\begin{table}[htbp]
\centering
\setlength{\tabcolsep}{4.0pt}
\begin{tabular}{lcccc}
\toprule
 & \multicolumn{2}{c}{Sound waves} & \multicolumn{2}{c}{Turbulence} \\
\cmidrule(lr){2-3}\cmidrule(lr){4-5}
Point & $f_{\rm peak}$ (Hz) & $h^2\Omega_{\rm peak}$ & $f_{\rm peak}$ (Hz) & $h^2\Omega_{\rm peak}$ \\
\midrule
\multicolumn{5}{l}{{Wall speed} $v_w=0.6$} \\
\addlinespace[2pt]
$P_1$ & $2.2$ & $3.5\times10^{-26}$ & $7.9$ & $1.0\times10^{-28}$ \\
$P_2$ & $5.0$ & $2.7\times10^{-27}$ & $18$ & $7.7\times10^{-30}$ \\
$P_3$ & $16$ & $1.9\times10^{-28}$ & $57$ & $4.7\times10^{-31}$ \\
$P_4$ & $19$ & $1.3\times10^{-28}$ & $69$ & $3.1\times10^{-31}$ \\
$P_5$ & $21$ & $2.8\times10^{-29}$ & $76$ & $7.3\times10^{-32}$ \\
$P_6$ & $26$ & $4.7\times10^{-29}$ & $94$ & $1.1\times10^{-31}$ \\
$P_7$ & $41$ & $2.2\times10^{-29}$ & $150$ & $5.3\times10^{-32}$ \\
\midrule
\multicolumn{5}{l}{{Wall speed} $v_w=0.95$} \\
\addlinespace[2pt]
$P_1$ & $1.4$ & $4.5\times10^{-28}$ & $5.0$ & $2.8\times10^{-30}$ \\
$P_2$ & $3.2$ & $3.9\times10^{-29}$ & $12$ & $2.1\times10^{-31}$ \\
$P_3$ & $9.8$ & $4.1\times10^{-30}$ & $36$ & $1.3\times10^{-32}$ \\
$P_4$ & $12$ & $2.7\times10^{-30}$ & $43$ & $8.3\times10^{-33}$ \\
$P_5$ & $13$ & $5.2\times10^{-31}$ & $48$ & $2.0\times10^{-33}$ \\
$P_6$ & $16$ & $1.1\times10^{-30}$ & $59$ & $3.1\times10^{-33}$ \\
$P_7$ & $26$ & $5.6\times10^{-31}$ & $93$ & $1.4\times10^{-33}$ \\
\bottomrule
\end{tabular}
\medskip
\caption{Peak frequency and dimensionless peak amplitude of each component at the seven completed points at the two wall speeds.  Each is the closed-form peak evaluated at the strength, inverse duration and nucleation temperature of Table~\ref{tab:completed-representative-thermal}.  The turbulent columns assume $\epsilon_{\rm turb}=0.10$ of Sec.~\ref{subsec:gw-spectra}.}
\label{tab:completed-benchmark-gw-peaks}
\end{table}

Table~\ref{tab:completed-benchmark-gw-peaks} gives the peak of each component at each point and wall speed.  The acoustic peaks confirm the bound: the largest of them, at $P_1$ and $v_w=0.6$, is $h^2\Omega_{\rm sw,peak}=3.5\times10^{-26}$, at least $6$ orders of magnitude below Eq.~\eqref{eq:gw-sw-bound}.  Across the benchmarks, raising the wall speed from $0.6$ to $0.95$ costs a factor of $40$ to $80$ in the acoustic peak amplitude.  The cost differs from point to point because three factors of Eq.~\eqref{eq:gw-components} change with the wall speed at once: $v_{\rm sh}$, the efficiency $\kappa_{\rm sw}$, which falls as a detonation becomes faster, and the lifetime factor $\Upsilon$.  The last interpolates smoothly between $\Upsilon\simeq H_*\tau_{\rm sw}$, an amplitude scaling as $(\beta/H_*)^{-2}$ for a short-lived source, and $\Upsilon\to1$, an amplitude scaling as $(\beta/H_*)^{-1}$ for a long-lived one.  The acoustic peak frequencies, from about $1$ to $40$~Hz, and the turbulent ones, up to about $150$~Hz, lie far above the millihertz band because of the large $\beta/H_*$ of Table~\ref{tab:completed-representative-thermal}.  Every tabulated sample of these spectra lies below the LISA, DECIGO and BBO sensitivities by at least $10$ orders of magnitude at the frequency where the gap is smallest.

The turbulent peaks sit a factor of a few hundred below the acoustic ones and about $3.6$ times higher in frequency.  Both ratios follow from Eqs.~\eqref{eq:gw-components}--\eqref{eq:gw-frequencies}.  At the peak frequencies $f\gg h_*$, so the last factor of $S_{\rm turb}$ in Eq.~\eqref{eq:gw-shapes} is close to $h_*/(8\pi f)$.  It moves the maximum of the turbulent shape from $x=9/2$ to $x=6/5$.  With $v_{\rm sh}=v_w$ at both wall speeds, the frequency ratio is then $(2.7/0.89)\times6/5\simeq3.6$ at every point.  The same factor gives the turbulent peak a second power of $v_w/\widetilde\beta$, through $h_*/f_{\rm turb}=0.61\,v_w/\widetilde\beta$, so the explicit factor $1/\widetilde\beta$ does not cancel in the ratio of the peaks.  With $K\equiv\kappa_{\rm sw}\alpha_\theta/(1+\alpha_\theta)$ and $\kappa_{\rm turb}=\epsilon_{\rm turb}\kappa_{\rm sw}$, the ratio is
\begin{equation}
 \frac{h^2\Omega_{\rm sw,peak}}{h^2\Omega_{\rm turb,peak}}
 \simeq4.0\,\frac{\Upsilon\,\widetilde\beta\,v_{\rm sh}}{v_w^2}\,
 \frac{K^{1/2}}{\epsilon_{\rm turb}^{3/2}}
 ~.
 \label{eq:gw-peak-ratio}
\end{equation}
For a short-lived source, $\Upsilon\simeq H_*\tau_{\rm sw}$ and Eq.~\eqref{eq:gw-acoustic-lifetime} give $\Upsilon\widetilde\beta=(8\pi)^{1/3}v_{\rm sh}/\overline U_f\propto K^{-1/2}$.  This removes both the duration and the fluid efficiency from Eq.~\eqref{eq:gw-peak-ratio}, and the ratio becomes $13.5\,\epsilon_{\rm turb}^{-3/2}\simeq425$ for $v_{\rm sh}=v_w$.  Since $\Upsilon\leq H_*\tau_{\rm sw}$, this is the largest value the ratio can take.  At $v_w=0.6$, where the sources are short-lived, the peaks of Table~\ref{tab:completed-benchmark-gw-peaks} give ratios from $350$ to $430$, close to it.  When the source lives longer, $\Upsilon$ falls below $H_*\tau_{\rm sw}$ and the ratio drops, to about $160$ to $190$ at $P_1$ and $P_2$ for $v_w=0.95$.  That the turbulent peak is the smaller at every point is a numerical result of Table~\ref{tab:completed-benchmark-gw-peaks}.  The turbulent estimate therefore adds nothing to the reach of the signal and is shown only to record the size of a contribution the effective potential does not determine.


The seven benchmark points demonstrate that the minimal type-II seesaw can complete an electroweak FOPT while satisfying the zero-temperature and electroweak-precision conditions.  Every completed transition found in the sampled subspace is weak, with $\Delta\varphi/T_c$ between $0.18$ and $0.44$, and fast, with $\beta/H_*$ between $4\times10^{4}$ and $1\times10^{6}$.  The first range spans only a factor of $2.4$ across points whose triplet VEVs span more than three orders of magnitude and whose doubly charged masses differ by a factor of two.  All seven points have $\lambda_3(Q_{\rm ref})=0$.  Recomputing $P_1$ and $P_7$ with $\lambda_3(Q_{\rm ref})=\pm0.5$, and $P_5$ with $+0.5$, shows that the transition is sensitive to $\lambda_3$ but does not become strong.  At $\lambda_3=-0.5$, $P_5$ violates $\lambda_2+\lambda_3\geq0$ at $Q_{\rm ref}$ and is not used.  A negative $\lambda_3$ lowers $T_c$ and weakens the transition: $\Delta\varphi/T_c$ falls from $0.44$ to $0.21$ at $P_1$ and from $0.18$ to $0.16$ at $P_7$.  At $P_7$, $\lambda_2+\lambda_3$ becomes negative below $76$~GeV and reaches $-0.019$ at $60$~GeV, but the point satisfies the conditions imposed at $Q_{\rm ref}$, and boundedness is not reimposed at intermediate scales in the present prescription.  Meanwhile, $\alpha_\theta$ changes by less than $15\%$, while $\beta/H_*$ rises to about $10^6$ at $P_1$ and $2\times10^6$ at $P_7$, which lowers the acoustic signal further.  At $\lambda_3=+0.5$, the direct FOPT from the symmetric phase disappears at $P_1$ and $P_7$, while the zero-temperature minimum moves away from $v$ at $P_5$.  At $\lambda_3=+0.5$, the deepest zero-temperature point of $P_1$ and $P_7$ already lies at the lower end of the running interval, $\varphi=60$~GeV, where the present prescription is not reliable.  Within this range, a nonzero $\lambda_3$ therefore changes or removes the transition, but nowhere produces a sufficiently strong one or an observable GW signal.

\section{Conclusions}
\label{sec:conclusions}

We studied electroweak thermal histories in the minimal type-II seesaw model with a one-loop, RG-improved finite-temperature effective potential, tracing every thermal history in the neutral field space $(\varphi,\delta)$ with CT version~2.0.7.

A broad survey was used to test the usual critical-temperature analysis.  Applied mechanically to every first-order degeneracy in the CT phase graphs, the cut $v_{\rm low}(T_c)/T_c>1$, with $v_{\rm low}$ the field on the low-temperature side of the degeneracy, selects a minority of the resolved first-order degeneracies, and all but one of them evolve to a zero-temperature vacuum other than the electroweak one.  The cut selects degeneracies whose parent phase is already far from symmetric.  Those with a positive latent heat have strengths many orders of magnitude above that of the strongest transition that completes, and inserting them into the standard GW formulae inflates the predicted signal by many orders of magnitude as well.

Seven benchmark parameter points pass every condition imposed in this work, including integrated nucleation, percolation and completion, the fit at $95\%$ CL of the oblique parameters, and the direct searches that apply at their triplet VEVs.  Each has a single direct FOPT from the symmetric phase to the electroweak vacuum.  At temperatures above the upper edge of the origin interval, between about $0.5$ and $1.4$~TeV at $\kappa=1$, the origin is no longer a local minimum.  The upper edge is produced by the running couplings through the near cancellation of the thermal and ring contributions.  The field-dependent scale adds a smaller negative curvature term and lowers the edge by about $100$~GeV.  These histories therefore assume that the plasma reaches the origin from higher temperatures.  All seven benchmarks assume $\lambda_3(Q_{\rm ref})=0$ for simplicity.  At three of them, setting $\lambda_3=+0.5$, or $-0.5$ where boundedness at $Q_{\rm ref}$ allows it, changes $\Delta\varphi/T_c$ by up to a factor of two, removes the transition, or moves the zero-temperature vacuum away from $v$, but nowhere makes the transition strong.

In the central prescription, the transitions are weak, with a field discontinuity $\Delta\varphi/T_c$ between $0.18$ and $0.44$ and a trace anomaly strength $\alpha_\theta$ of a few times $10^{-5}$, fast, with an inverse duration in Hubble units $\beta/H_*$ between $4\times10^{4}$ and $1\times10^{6}$, and almost free of supercooling, $T_n$ lying below $T_c$ by at most $0.6\%$.  A low and narrow barrier is consistent with the small field discontinuity, the small supercooling and the short duration, while the small strength is a separate feature of these points.  That strength is a small residue of much larger one-loop terms.  Its magnitude is known only to within one to two orders of magnitude, and its sign is not stable under the radial rematching diagnostic.  The screening of the zero modes by the triplet states, which are about twice as heavy as $T_c$, is taken in the massless approximation.  The broken-phase field at nucleation, $\xi_n=\varphi(T_n)/T_n$ between $0.19$ and $0.46$, falls short of the value of about $1$ that the sphaleron washout estimate requires.  Within this estimate, which uses one sphaleron shape factor for all the benchmark points, baryon number preservation is not supported at any of them.  The point with the strongest transition is also the one closest to the edge of the electroweak fit.

The seven benchmarks found in this study lie above the triplet VEV where the same-sign dilepton and same-sign diboson widths of the doubly charged Higgs boson are equal, by more than an order of magnitude.  Therefore, the prompt-dilepton limits of ATLAS at the LHC in the higher mass regime do not apply to them.  At the six points with $\lambda_5>0$, the doubly charged state decays to same-sign $WW$ pairs and lies above the $350$~GeV ATLAS pair production limit in that channel.  At $P_5$, it decays mainly through the cascade process $H^{++}\to H^+W^{+*}$.

The renormalization scale was varied at all seven points over $0.50\leq\kappa\leq1.50$, and the comparison with the unimproved potential shows why the RG improvement is needed.  With the improvement, the running of the couplings compensates the explicit scale dependence of the one-loop terms.  Every point then keeps its transition throughout $0.50\leq\kappa\leq1.00$.  What remains is a change of $\Delta\varphi/T_c$ by a factor of a few, the residual scale dependence of this one-loop prescription.  Without the improvement, the couplings stay at $Q_{\rm ref}$ while the logarithms are taken at the varied scale, so nothing compensates for that dependence and the two potentials differ already at one loop.  Without the improvement, none of the seven benchmarks has a FOPT into a stable vacuum at any scale choice.

The GW signal of the studied benchmark points is far below the reach of LISA, DECIGO and BBO.  In the central prescription, the largest computed acoustic peak is $3.5\times10^{-26}$.  Replacing $\alpha_\theta$ with any of the values obtained when the running-vacuum term or the radial-rematching term is removed leaves every computed peak more than $5$ orders of magnitude below the deepest point of the nearest sensitivity curve.

\section*{Acknowledgments}
\label{sec:acknowledgments}

This work was supported in part by the National Science and Technology Council under Grant No.~NSTC114-2112-M-002-020-MY3.  AI-assisted tools were used in writing the computer codes, producing the figures, and editing the language of the manuscript.  The author takes full responsibility for the content of this paper.

\end{document}